\documentclass[12pt]{article}
\usepackage{amsfonts}
\usepackage{amssymb,amsmath,amsthm,amsfonts}
\usepackage{graphicx}
\theoremstyle{plain}
\newtheorem{theorem}{Theorem}[section]
\newtheorem{lemma}{Lemma}[section]
\newtheorem{proposition}{Proposition}[section]

\theoremstyle{definition}

\newcommand{\X}{{\mathcal X}}

\begin{document}

\title{{\bf Inside Trading, Public Disclosure and Imperfect Competition\thanks{For useful discussions, we thank Deqing Zhou.
We also would like to thank Jiaan Yan, Jianming Xia and other
seminar participants at the Institute of Applied Mathematics,
 Academy of Mathematics and Systems Science,
Chinese Academy of Sciences. The first named author is grateful for
financial support for National Natural Science Foundation of China
(No.10721101) and China's National 973 Project (No.2006CB805900).
The second author  also would like to thank Hengfu Zou and other
seminar participants at Central University of Finance and Economics.
}}}
\author{Fuzhou Gong$^1$ \quad
  Hong Liu$^{2,3}$ \thanks{E-mail: mathhongliu@gmail.com.}
  \\ {\small  $^{1}$ Institute of Applied mathematics, Academy of
Mathematics and Systems Science, }
\\ {\small Chinese Academy of Sciences,
Beijing 100190, P.R. China.}
\\ {\small  $^{2}$MOE Key Laboratory of Applied Statistics, School of Mathematics and Statistics,}
\\
{\small  Northeast Normal University, Changchun, 130024, China}
\\ {\small  $^{3}$China Economics and Management Academy, and CIAS,}
\\ {\small Central University of Finance and Economics, Beijing, 100081, P.R. China.}
\date{}}
\maketitle

\begin{abstract}
In this paper, we present a multi-period trading model in the style
of Kyle (1985)'s inside trading model, by assuming that there are at
least two insiders in the market with long-lived private
information, under the requirement that each insider publicly
discloses his stock trades after the fact. Based on this model, we
study the influences of ``public disclosure'' and ``competition
among insiders'' on the trading behaviors of insiders. We find that
the ``competition among insiders'' leads to higher effective price
and lower insiders' profits, and the ``public disclosure'' makes
each insider play a mixed strategy in every round except the last
one. An interesting find is that  as the total number of auctions
goes to infinity, the market depth and the trading intensity at the
first auction are all constants with the requirement of ``public
disclosure'',  while  the market depth at the first auction goes to
zero and the trading intensity of the first period goes to infinity
without the requirement of ``public disclosure''.
 Moreover, we give the exact
 speed of the revelation of the private information, and show that
all information is revealed immediately and the market depth goes to
infinity immediately as trading happens infinitely frequently.

 \vspace{0.3cm}


\vspace{0.3cm}

\end{abstract}

 \setcounter{equation}{0}
In the field of mathematical  finance,
 lots of  famous models are based on the assumption that all
 traders in the market have the same information and the same
 expectation. However, many financial and commodity markets can be characterized by a number of insiders,
  each with the same or different information. How will an insider trade in the market?  How valuable is the private information to an
 insider? Moreover, how efficient are financial markets at incorporating information?
 These interesting questions have generated a large body of research,
 both theoretical and empirical.

In his pioneering and insightful paper, Kyle (1985)  proposes a
model pricing a risky asset in the presence of new private
information, and obtains a unique linear equilibrium  by assuming
that the ex-ante asset value is normally distributed and the price
is a linear function of the aggregated market order. The
monopolistic insider, in order to maximize his conditional profit,
will trade in a recursive manner in discrete model, while in
continuous time case when the time interval goes to zero, the
private information is incorporated into market price at a constant
speed, and the market depth is constant over time. Based on Kyle
(1985)'s model, Holden and Subrahmanyam (1992) give a model in which
there are more than two informed traders in the market, and find
that informed traders will trade aggressively and the market depth
becomes extremely large almost immediately. The only difference
between the two papers
 is that, Holden and Subrahmanyam (1992) assumes that at least two insiders have
 the same information, and Kyle (1985) assumes that  there is only one monopoly insider in the market.
 The intuition lies in that, there is a trade-off of two effects between
information for one agent and that for others in multiple agents
decision games. Better information may improve an agent's decision,
but this may also cause other agents' decisions to strategically
shift, and which in turn has an impact on the original agent's
decisions. So, in Holden and Subrahmanyam (1992)'s model, each
trader tries to beat the
 others to the market, with the result that their information is
 revealed almost immediately. Comparing these two papers we can get
 the conclusion that, the number of insiders  has a big
 influence on the market structure.

Now, Kyle (1985)'s  model has been widely used to analyze
microstructure of financial market and the value of information, and
has elicited a large body of literature. For example, Foster and
Viswanathan (1996) also consider a market with multiple competing
insiders, but with the  assumption that each informed trader's
initial information is  a noisy estimate of the long-term value of
the asset, and the different signals received by informed traders
can have a variety of correlation structures.  Back (1992)
formalizes and extends Kyle (1985)'s model by showing the existence
of a unique equilibrium beyond the Gaussian-linear framework.
Remarkable, when the asset value has a log-normal distribution, the
price process becomes a geometric Brownian motion, as  usually
assumed in finance. Admati and Pfleiderer (1988) analyze a strategic
dynamic market order model. Their model is essentially a dynamic
repetition of a generalized version of the static model in Kyle
(1985). However, their focus is on intraday price and volume
patterns. They attempt to explain the U-shape of the trading volume
and price changes, that is, the abnormal high trading volume and
return variability at the beginning and at the end of a trading day.
When there exists public information, Luo (2001) extends Kyle
(1985)'s model by showing that,
 the monopoly insider put a
negative weight on the public information in formulating his optimal
strategy.  Recently, Caldentey and Stacchetti (2010) study the
extended Kyle (1985)'s model by assuming that an insider
continuously observes a signal that trackes the evolution of asset's
fundamental value and the value of the asset is  publicly revealed
at a random time.  Moreover, Chau and Vayanos (2008), Fishman and
Hagerty (1992) and  Rochet and Vila (1994) etc. have used the
variants of Kyle's model to analyze and to explain real financial
phenomena.

In addition to the literature mentioned above,  Huddart Hughes and
Levine (2001) present an insider's equilibrium trading strategy in a
multi-period rational expectations framework based on Kyle (1985),
given the requirement that the  insider must publicly disclose his
stock trades after the fact. Just as Hudart, Hughes and Levine
(2001) say, under US securities laws,  insiders, such as officers
directors and beneficial owners of five percent or more of equity
securities associated with a firm,  must report to the Securities
and Exchange Commission (SEC) trades they make in the stock of that
firm. These reports are filed after the trade is completed, and are
publicly available immediately upon receipt by the SEC. So it is
necessary and interesting to study the effect of ``public
disclosure'' to the dynamic and continuous trading strategy of
informed traders. In their paper,  they provide a solution to a
discrete time analog of Kyle (1985)'s rational expectations trading
model, where an insider endowed with long-lived private information
must disclose the quantity he trades at the end of each round of
trading. They also find that the insider garbles the information
conveyed by his trade by playing a mixed strategy in every round
except the last one. Nevertheless, the public disclosure of the
insider's trades accelerates not only the price discovery process
but also the trading intensity of the insider  comparing with Kyle
(1985)'s model.

 Inspired by Holden and Subrahmanyam (1992) and  Huddart Hughes and Levine
 (2001),  we
consider a  model  in which there are at least two informed traders
in the
 market,  with the requirement that  insiders  publicly disclose their
 stock trades after the fact. In this
paper, we consider the existence and the uniqueness of  insiders'
equilibrium trading strategy in a multi-period rational expectation
framework and give the analysis of the equilibrium. In equilibrium,
the ``trade public disclosure'' and  ``competition among insiders''
have great effects on each  insider's trading intensity, the market
depth and the effectiveness of the price. Comparing with
 Holden and Subrahmanyan (1992)'s model,   each  insider plays a mixed
strategy in every round except the last one, because of the
existence of ``public disclosure'', which also leads to the
accelerated price discovery  and  higher market depths. Furthermore,
in the sequential auction equilibrium, market depths become infinite
and all private information is revealed immediately when the time
interval between auctions approaches zero. The speed of the
revelation is faster than that of Holden and Subrahmanyan (1992),
and we give the exact speed of the revelation. Although market depth
becomes infinite immediately, the initial market depth
 in our model is a constant, as the total number
of auction goes to infinity. This is very different from that of
Holden and Subrahmanyan (1992), where as the number of trading
rounds per unit time becomes larger, the market depth  in the
initial period becomes larger. In Holden and Subrahmanyan (1992),
the adverse selection  is high in the earlier periods because the
information content of the order flow is high, and negligible in the
later periods because market makers have very little to fear from
traders that have already exploited most of their informational
advantage. As the number of auctions is increased, trading becomes
more concentrated at earlier auctions and hence there is greater
adverse selection at these auctions. But under the requirement of
``public disclosure'', market makers know that insiders may disclose
their trading and may choose a mixed strategy to dissimulate their
information,  the adverse selection is the same at the initial
auction even if the number of auctions increases.

Related to Huddart Hughes and Levine
 (2001),  ``competition among insiders'' leads to more effectiveness of the
 price and lower profits of insiders. In particular, in the two-period trading, the market depth
 of the first auction in
our model is smaller
 than Hudart, Hughes
 and Levine (2001)'s when the number of insiders
 is less than $5$, however, the opposite conclusion holds if the number of insiders is more than $5$. That is to say,
 whether market makers set the marginal cost of the first period trades lower or higher depends on the number of insiders.
 If the number of insiders is big enough (bigger than $5$), the adverse selection becomes small.

Moreover, we give the near-continuous trading results by starting
with discrete time and then taking the limit, rather than
formulating the model directly in continuous time. When the
intervals between auctions becomes uniformly small, we give the
expression of the error variance of the price and the depth of the
market depth at any positive time. We find that  the speed of
insiders' information incorporated into the price is  a linear
function of the time $t$  when there is one insider in the market.
However, if there are more than two insiders, then all information
is revealed immediately and the market depth goes to infinity
immediately as the number of auctions goes to infinite where the
market is strong-form efficient. An interesting find is that  as the
total number of auction goes to infinity, the market depth and the
trading intensity at the first auction are all constants while
Holden and Subrahmanyam (1992) find that the market depth at the
first auction goes to zero and the trading intensity of the first
period goes to infinity. The intuition for the contrast is that
``public disclosure'' enable  less adverse selection in the market.

Furthermore,  from  Section $2.2$ to Section $2.4$ in  Gong and Zhou
(2010), we know that there exists an inconsistency between
``constant pricing rule"  in the definition of ``linear equilibrium"
and the implication of ``market efficiency" in Kyle (1985). In order
to modify the inconsistency, Gong and Zhou (2010) consider three
different models according to the insider's attitudes regarding to
risk, in which the basic assumptions of the model $2$ are the same
as that of Kyle (1985). The analysis of model $2$ and  Theorem $4$
in Gong and Zhou (2010) imply that the equilibrium results of model
$2$ are the same as that of Kyle (1985),  as trading happens
infinitely frequently. Therefore, using the same method to modify
the inconsistency as Gong and Zhou (2010) do,  it is possible that
we can get the similar results as our paper when the trading happens
infinitely frequently. But through preliminary research, we know
that the mixed strategy in game theory may be involved in the
modified model, which leads to very complex calculations. We will
consider this problem in other papers, and the purpose of this paper
is to give the extension of Holden and Subrahmanyam (1992) and
Hudart, Hughes
 and Levine (2001) using the framework of Kyle (1985) so as to
 compare with the modified model.

 The rest of this
paper is structured as follows. Section $1$ presents the  analysis
of insiders' strategy and gives the two-period equilibrium of Holden
and Subrahmanyam (1992)'s model and that of our model with public
disclosure requirement. Section $2$ determines the unique linear
Nash equilibrium in multi-period rational expectations framework and
gives the analysis of the equilibrium.  Section $3$ considers the
properties of the equilibrium when the trading is carried out
continuously. In particular, the exact convergence rate of error
variance is given in Section $3$. Section $4$ concludes  the paper
and Section $5$ is the Appendix, which contains the proof of the
theorem and
 some necessary propositions.

\section{Analysis}
\subsection{Two-period Holden and Subrahmanyam (1992)'s Model}
\setcounter{equation}{0} \quad \quad In order to analysis the
equilibrium strategy that every insider chooses in our model we use
the same structure as  Hudart, Hughes and Levine (2001), in which
the equilibrium strategy of a monopolist insider is given. Our
starting point is a two-period discrete-time model based on Holden
and Subrahmanyam (1992), with several informed traders who obtain a
signal about the value of a risky asset. Consider a standard Holden
and Subrahmanyam (1992) two-period model, in which there is one
risky asset with a liquidation value, $v$, that is normally
distributed with prior mean $p_0$ and prior variance $\Sigma_0$. Let
$M$ denote the number of informed traders, who are indexed by
$i=1,\cdots M$. Index the periods by $n\in\{1,2\}$. Let $x_n$ and
 $x_{in}$ denote the total order by all informed traders and
  the individual order by the $i$th informed trader at the
$n$th auction, respectively. Let $\pi_{i1}$  and $\pi_{i2}$ denote
the portion of insider $i$'s total profits directly attributable to
his period $1$ and period $2$ trade, respectively.
  Each informed trader
learns $v$ at the start of the first period and places an order to
buy or sell $x_{in}$ shares of the risky asset at the start of
period $n (n\in\{1,2\})$. Market makers receive these orders along
with those of liquidity traders whose exogenously-generated demands,
$\mu_n$, are normally distributed with mean $0$ and variance
$\sigma_{\mu}^2$. Assume $v$, $\mu_1$ and $\mu_2$ are mutually
independent. At each auction,  market makers
 observe only the total order flow $y_n=x_n+\mu_n$, but not each of them,  and set the price $p_n$,
 the price at the $n$th auction. We assume that there is no
discount across the two periods, that is, the interest rate is
normalized to zero.

Equilibrium is defined by the  market efficiency condition that
$p_n$ equals to the expected value of $v$ conditional on the
information available to the market makers at the $n$th auction, by
a profit maximization condition that each informed trader selects
the optimal strategy conditional on his conjectures and his
information at each auction, and by a condition that all conjectures
are correct.

Let the insider $i$'s trading strategy and market makers' pricing
rule be sets of real-valued functions $X_i=\{X_{i1},X_{i2}\}$ and
$P=\{P_1,P_2\} $ such that, given an initial price $p_0$,
$$x_{i1}=X_{i1}(p_0,v), ~~~~x_{i2}=X_{i2}(P_{1},v),
~~~~p_1=P_1(y_1), ~~~~p_2=P_2(y_1,y_2).
$$
It is apparent that $\pi_{in}(x_{in},p_n)=x_{in}(v-p_n)$. A sub-game
perfect equilibrium is defined by $X_i$ and $P$ such that, for all
$n\in \{1,2\}$, $i=1, \cdots ,M$,  and for any other strategy
$\hat{X}_i=\{\hat{X}_{i1},\hat{X}_{i2}\}$;
$$E[\pi_{i1}(x_{i1},p_1)+\pi_{i2}(x_{i2},p_2)|p_0,v]\geq E[\pi_{i1}(\hat{x}_{i1},p_1)+\pi_{i2}(\hat{x}_{i2},p_2)|p_0,v],$$
$$E[\pi_{i2}(x_{i2},p_2)|p_0,p_1,v]\geq E[\pi_{i2}(\hat{x}_{i2},p_2)|p_0,p_1,v];$$
and
$$p_1=E(v|y_1);~~~~~p_2=E(v|y_1,y_2).$$
Define $\Sigma_1=Var(v|y_1)$ and
 $\Sigma_2=Var(v|y_1,y_2)$, which are measures of the
 informativeness of prices.
The proposition below is based on a special case of Proposition $1$
in  Holden and Subrahmanyam (1992).
\begin{proposition}\label{pro1}Given no public disclosure of
insider trades, a two-period subgame perfect linear equilibrium
exists in which
$$x_{in}=\beta_n(v-p_{n-1}),~~~~n\in\{1,2\},$$
$$p_n=p_{n-1}+\lambda_ny_n,~~~~n\in\{1,2\},$$
$$\lambda_1
=\frac{\sqrt{M(M+1)^2[(M+1)^2-2k]\Sigma_0}}{[(M+1)^3-2kM]\sigma_{\mu}},~~~~\lambda_2=
\frac{1}{\sigma_{\mu}}\sqrt{\frac{M\Sigma_0}{(M+1)^3-2kM}},$$
$$\beta_1=\frac{(M+1)^2-2k}{\lambda_1[(M+1)^3-2kM]},~~~~~\beta_2=\frac{1}{\lambda_2(M+1)},$$
$$\Sigma_1=\frac{(M+1)^2\Sigma_0}{(M+1)^3-2kM},~~~~\Sigma_2=\frac{1}{M+1}\Sigma_1=\frac{(M+1)\Sigma_0}{(M+1)^3-2kM},$$
where  $ k\doteq \frac{\lambda_1}{\lambda_2}$, satisfies the
equation
$$2Mk^3-(M+1)^3k^2-2(M+1)^2k+(M+1)^4=0,$$
and $0<k<\frac{(M+1)^2}{2}$.
\end{proposition}
\begin{proof}
See the Appendix.
\end{proof}
Note that the above proposition provides a benchmark against which
to compare an equilibrium for the case where  insiders' trades in
the first period are publicly disclosed.

\subsection{Two-period model with public disclosure of
          insider trades} \setcounter{equation}{0}

\quad \quad  Using  the same method  by Hudart, Hughes
 and Levine (2001), we can know that no invertible trading strategy
 can be part of an equilibrium in this case, and
 we  show that an equilibrium exists, in which
insiders' first-period trade consists of an information-based linear
component, $\beta_{i1}$, and a noise component, $z_1$, where $z_1$
is normally distributed with mean $0$ and variance $\sigma_{z_1}^2$,
and independently with $v$ and $\mu_1$. For  market makers, public
disclosure of $x_1$ allows them to update their beliefs from those
formed on a basis of the first period order flow. In particular, let
$p_1^*=p_0+\gamma_1x_1$ be the expected value of $v$ given $x_1$ and
$y_1$. Thus, $E(v|x_1,y_1,p_1)=E(v|x_1)$. In turn, $p_1^*$ replaces
$p_1$ in the second period price $p_2=p_1^*+\lambda_2y_2$. This
pricing rule is accepted by  insiders because the cost of the
information that can bring profit to them is zero. Using the
symmetry of strategy for each insider, it is straightforward to show
that, given linear pricing rule by  market makers, the only possible
equilibrium among  insiders is one in which they choose the same
strategy. Accordingly, we can take one insider to analysis. Here,we
take the insider $i$.

 Applying the principal of
backward induction, we can write the insider $i$'s second period
optimization problem for given $x_{11}, x_{21},\cdots, x_{M1} $ and
$p_1^*$ as $x_{i2}\in \arg\max_x E[x(v-p_2)|x_{11}, x_{21},\cdots,
x_{M1} ,p_1^*,v]$, where
\begin{equation}{\label{eq3.1}}
\begin{aligned}
&E[x(v-p_2)|x_{11}, x_{21},\cdots, x_{M1},p_1^*,v]\\
=&[v-p_1^*-\lambda_2x-\lambda_2E(\sum_{j\neq i}x_{j2}|x_{11},
x_{21},\cdots, x_{M1},p_1^*,v)]x.
\end{aligned}
\end{equation}
  Solve the above maximization problem
yields the maximizing value of $x$, which we denote as $x_{i2}$,
\begin{equation}\label{eq3.3}
x_{i2}=\frac{1}{2\lambda_2}(v-p_1^*)-\frac{1}{2}E(\sum_{j\neq
i}x_{j2}|x_{11}, x_{21},\cdots, x_{M1},p_1^*,v).
\end{equation}
By the theory of solving linear equations, we know
$x_{12}=x_{22}=x_{32}=\cdots=x_{M2}$. So
\begin{equation}\label{eq3.4}x_{i2}=\frac{1}{(M+1)\lambda_2}(v-p_1^*),
\end{equation}
and
\begin{equation}\label{eq3.5}E[\pi_{i2}(p_1^*,v)|x_{11},
x_{21},\cdots,
x_{M1},p_1^*,v]=\frac{1}{(M+1)^2\lambda_2}(v-p_1^*)^2.
\end{equation}
The second order condition is $\lambda_2>0$.

Stepping back to the insider $i$'s first period optimization
problem, we have
$$x_{i1}\in \arg \max _x E[x(v-p_1)+\pi_{i2}(p_1^*,v)|v],$$
where
\begin{equation}{\label{eq3.6}}
\begin{aligned}
&E[x(v-p_1)+\pi_{i2}(p_1^*,v)|v]
=x[v-p_0-\lambda_1x-\lambda_1E(\sum_{j\neq i}x_{j1}|v)]\\
&+\frac{1}{(M+1)^2\lambda_2}(v-p_0-\gamma_1x-\gamma_1E(\sum_{j\neq
i}x_{j1}|v))^2,
\end{aligned}
\end{equation}
 the first order condition implies
\begin{equation}{\label{eq3.7}}
\begin{aligned}
&v-p_0-2\lambda_1x_{i1}-\lambda_1E(\sum_{j\neq i}x_{j1}|v)\\
&+\frac{-2\gamma_1}{(M+1)^2\lambda_2}(v-p_0-\gamma_1x_{11}-\gamma_1E(\sum_{j\neq
i}x_{j1}|v))=0.
\end{aligned}
\end{equation}
Also by the theory of solving linear equations, we have
$x_{11}=x_{21}=x_{31}=\cdots=x_{M1}$, and
\begin{equation}{\label{eq3.8}}\left(1-\frac{2\gamma_1}{(M+1)^2\lambda_2}\right)(v-p_0)+\left(\frac{2M\gamma_1^2}
{(M+1)^2\lambda_2}-(M+1)\lambda_1\right)x_{i1}=0.
\end{equation}
The second order condition is
$$\frac{2M\gamma_1^2}
{(M+1)^2\lambda_2}-(M+1)\lambda_1\leq 0.$$

If our proposed mixed trading strategy,
$x_{i1}=\beta_{i1}(v-p_0)+z_{i1},$ $z_{i1}\sim N(0,\sigma_{z_1}^2),$
is to hold in equilibrium, by (\ref{eq3.8}) we can get that
$\lambda_1$,$\lambda_2$, and $\gamma_1$ satisfy
$\lambda_1>0,\lambda_2>0$, and
\begin{equation}{\label{eq3.9}}1-\frac{2\gamma_1}{(M+1)^2\lambda_2}=0,
\end{equation}
\begin{equation}{\label{eq3.10}}\frac{2M\gamma_1^2}
{(M+1)^2\lambda_2}-(M+1)\lambda_1=0.
\end{equation}
Combining (\ref{eq3.9}) and (\ref{eq3.10}) we have
\begin{equation}{\label{eq3.11}}\lambda_2=\frac{2\gamma_1}{(M+1)^2},
\end{equation}
\begin{equation}{\label{eq3.12}}\lambda_1=\frac{M\gamma_1}{M+1}.
\end{equation}
The market efficient condition implies
\begin{equation}{\label{eq3.13}}p_1=E(v|y_1)=p_0+\lambda_1y_1,~~p_1^*=E(v|x_1)=p_0+\gamma_1
x_1,~~p_2=E(v|y_2,p_1^*)=p_1^*+\lambda_2y_2,
\end{equation}
where $y_1=\sum_{j=1}^Mx_{i1}+\mu_1$, $x_1=\sum_{j=1}^Mx_{i1}$,
$y_2=\sum_{j=1}^Mx_{i2}+\mu_2.$ By the projection theorem for normal
random variables, we get
\begin{equation}{\label{eq3.14}}\lambda_1=\frac{Cov(v,y_1)}{Var(y_1)}=
\frac{\sum_{j=1}^M\beta_{i1}\Sigma_0}{(\sum_{j=1}^M\beta_{i1})^2\Sigma_0+M\sigma_{z_1}^2+\sigma_{\mu}^2},
\end{equation}
\begin{equation}{\label{eq3.15}}\gamma_1=\frac{Cov(v,x_1)}{Var(x_1)}=
\frac{\sum_{j=1}^M\beta_{i1}\Sigma_0}{(\sum_{j=1}^M\beta_{i1})^2\Sigma_0+M\sigma_{z_1}^2}.
\end{equation}
 In order to get the expression of $\lambda_2$,  consider
 $$p_2-p_1^*=E(v-p_1^*|y_2)=E(v-p_1^*|\sum_{j=1}^M\beta_{i2}(v-p_1^*)+\mu_2),$$
Using the projection theorem for normal random variables again,   we
have
\begin{equation}\label{eq3.16}\lambda_2=\frac{\sum_{j=1}^M\beta_{i2}\Sigma_1}{(\sum_{j=1}^M\beta_{12})^2
\Sigma_1+\sigma_{\mu}^2},
\end{equation}
where
\begin{equation}\label{eq3.17}
\begin{aligned}
\Sigma_1&=Var(v|x_1)=\Sigma_0-\gamma_1^2((\sum_{j=1}^M\beta_{i1})^2\Sigma_0+M\sigma_{z_1}^2).
\end{aligned}
\end{equation}
$ (\ref{eq3.12})$,  $(\ref{eq3.14}) $ and  $(\ref{eq3.15})$
 imply
\begin{equation}\label{eq3.18}M(\beta_{11}^2\Sigma_0+\sigma_{z_1}^2)=\sigma_{\mu}^2,~~~~
\lambda_1=\frac{M\beta_{11}\Sigma_0}{(1+M)\sigma_{\mu}^2}.
\end{equation}
By $(\ref{eq3.11})$, $(\ref{eq3.17})$ and $(\ref{eq3.18})$  we  get
\begin{equation}\label{eq3.19}\Sigma_1=\Sigma_0-\frac{(M+1)^4}{4}\lambda_2^2M\sigma_{\mu}^2,
\end{equation}
and $(\ref{eq3.4})$ implies
\begin{equation}\label{eq3.20}\beta_{12}=\frac{1}{(M+1)\lambda_2}.
\end{equation}
Substituting this value for $\beta_2$ into  $(\ref{eq3.16})$ yields
$$\lambda_2=\frac{\sqrt{M\Sigma_1}}{(M+1)\sigma_{\mu}}.$$
In turn, $(\ref{eq3.19})$ reduces to
\begin{equation}\label{eq3.21}\Sigma_1=\Sigma_0-\frac{M^2(M+1)^2}{4}\Sigma_1.
\end{equation}
So it is easy to get that $\Sigma_1=\frac{4}{4+M^2(M+1)^2}\Sigma_0$.

According to the above analysis, we get the following proposition.
\begin{proposition}\label{pro2} There exists a unique linear equilibrium in the
two-period setting with public disclosure of insider trades, in
which there are constants $\beta_{i1}$, $\beta_{i2}$, $\lambda_1$,
$\lambda_2$, $\gamma_1$ and $\Sigma_1$, characterized by the
following:
$$x_{i1}=\beta_{i1}(v-p_0)+z_1, ~~~~x_{i2}=\beta_{i2}(v-p_1^*),$$
$$p_1=p_0+\lambda_1y_1,~~~p_2=p_1^*+\lambda_2y_2,~~~p_1^*=p_0+\gamma_1x_1,$$
$$\Sigma_1=Var(v|y_1,p_1)=Var(v|x_1),$$
for all informed traders $i=1,\cdots,M$. The constants $\beta_{i1}$,
$\beta_{i2}$, $\lambda_1$, $\lambda_2$, $\gamma_1$ and $\Sigma_1$,
satisfy:
$$\lambda_1
=\frac{M}{\sigma_{\mu}}\sqrt{\frac{M\Sigma_0}{4+M^2(M+1)^2}},~~~\lambda_2=
\frac{2}{(M+1)\sigma_{\mu}}\sqrt{\frac{M\Sigma_0}{4+M^2(M+1)^2}},$$
$$$$
$$\gamma_1=\frac{(M+1)\lambda_1}{M}=\frac{(M+1)}{\sigma_{\mu}}\sqrt{\frac{M\Sigma_0}{4+M^2(M+1)^2}},$$
$$\beta_{i1}=(M+1)\sigma_{\mu}\sqrt{\frac{M}{[4+M^2(M+1)^2]\Sigma_0}},~~~
\beta_{i2}=\frac{\sigma_{\mu}}{2}\sqrt{\frac{4+M^2(M+1)^2}{M\Sigma_0}},$$
$$$$
$$\sigma_{z_1}^2=\sigma_{\mu}^2-M\beta_{i1}^2\Sigma_0=\frac{4\sigma_{\mu}^2}{[4+M^2(M+1)^2]},$$
$$\Sigma_1=\frac{4\Sigma_0}{4+M^2(M+1)^2},$$
$$E[\pi_{i1}]=E[x_{i1}(v-p_1)]=\frac{[4+M^3(M+1)^2]\sqrt{M\Sigma_0}
\sigma_{\mu}}{[4+M^2(M+1)^2]^{\frac{3}{2}}},$$
$$E[\pi_{i2}]=\frac{1}{(M+1)^2\lambda_2}\Sigma_1=\frac{2\sqrt{\Sigma_0}\sigma_{\mu}}{(M+1)\sqrt{M[[4+M^2(M+1)^2]}}.$$
\end{proposition}

{\bf Analysis of the equilibrium:} When $M=1$, the above proposition
is consistent with
  Proposition $2$ in  Hudart, Hughes
 and Levine (2001). Proposition \ref{pro2}  implies
that $\lambda_1=\frac{M(M+1)}{2}\lambda_2$, so $\lambda_1>\lambda_2$
when there are more than two insiders in the market, that is to say
the depth of the market of the first period becomes smaller than
that of the second period as the number of insiders increases. That
is because the adverse selection  is high in first period as the
number of insiders increases. Proposition \ref{pro2} also implies
$\beta_{i1}<\beta_{i2}$, $i=1,2\cdots, M$ which means that the
trading intensity of the first period is lower than the second
period. This is consistent with the absence of a concern for the
effect of trading in the last period on future expected profits.
Moreover, insiders set
$Var(X_1)=M^2\beta_{i1}^2\Sigma_0+M^2\sigma_{z_1}^2=M\sigma_{\mu}^2$
($i=1,2\cdots, M$) to disguise their trading  by liquidity traders.

The same exogenous parameters imply different values for the
endogenous parameters depending not only on whether insiders must
disclose their traders after the fact but also on the number of
insiders. To distinguish the values, we add an upper bar to the
endogenous parameters in the case of Hudart, Hughes
 and Levine (2001), and a tilde to
the endogenous parameters in the case of Holden and Subrahmanyam
(1992). The next proposition compares the endogenous parameters
across the case of one insider and at least two insiders.

\begin{proposition}\label{pro3}In the two-period setting, the endogenous
parameters across the case of our model and Hudart, Hughes
 and Levine (2001)'s model satisfy:
  $$\lambda_1>\bar{\lambda}_1,~~~~(2\leq M \leq 5)$$
 $$\lambda_1<\bar{\lambda}_1,~~~~(M>5)$$
 and for all $M$, and $i=1,2\cdots, M$
 $$\lambda_2<\bar{\lambda}_2, ~~~~ \beta_{i1}<\bar{\beta}_1,$$
$$\beta_{i2}>\bar{\beta}_2, ~~~~\Sigma_1<\bar{\Sigma}_1,$$
$$E(\pi_{i1})<E(\bar{\pi}_{1}),~~~~
E(\pi_{i2})<E(\bar{\pi}_{2}).$$
\end{proposition}
\begin{proof}See the Appendix.
\end{proof}

The following proposition compares the endogenous parameters across
the case of disclosure  and no disclosure. For convenience, we only
analysis the case of $M=2$.

\begin{proposition}\label{pro4}In the two-period setting, the endogenous
parameters across the case of our model and Holden and Subrahmanyam
(1992)'s model satisfy:
$$\lambda_1<\tilde{\lambda}_1,~~\lambda_2<\tilde{\lambda}_2,$$
$$  \beta_{i1}>\tilde{\beta}_1,~~\beta_{i2}>\tilde{\beta}_2,$$
$$ \Sigma_1<\tilde{\Sigma}_1,$$
for all  $i=1,2.$
\end{proposition}
\begin{proof}See the Appendix.
\end{proof}

 Proposition $\ref{pro3}$ implies that, in the first auction, the market depth of our model is smaller
 than Hudart, Hughes
 and Levine (2001)'s when the number of insiders
 is less than $5$, and  the case is inverse if the number of insiders is more than $5$. That is to say,
 whether market makers set the marginal cost of the first period trades lower or higher depends on the number of insiders.
 If the number of insiders is bigger enough (bigger than $5$), the adverse selection becomes low.   Proposition $\ref{pro3}$
  also implies that $\beta_{i1}<\bar{\beta_1}$, i.e.,  every
 insider trades less intensely
 than Hudart, Hughes
 and Levine (2001)'s model because of competition.
 It is just the competition that makes all the insiders want
 to reveal their private information in the first trading period, resulting in a
more price informative, i.e. $\Sigma_1<\bar{\Sigma}_1$. According to
Proposition $\ref{pro4}$ we  know that, if  there are two insiders
in the model, then  market makers set the market depth of first
period trades higher with public disclosure than without, i.e.
$1/\lambda_1>1/\tilde{\lambda}_1$, under the rational conjecture
that some of  insiders' trades are randomly generated in the former
case. Every insider trades more intensely with public disclosure
than without in the first period, i.e. $\beta_{i1}>\tilde{\beta}_1$.
The measure of the informativeness of prices is smaller with public
disclosure than without, i.e. $\Sigma_1<\tilde{\Sigma}_1$. It  means
that  the price reflects more information than without because of
``public disclosure''.

Now we analysis the second period strategy.
  Because
more information is incorporated into the price in the first period,
every insider trades more intensely than that of Hudart, Hughes
 and Levine (2001)'s and Holden and Subrahmanyam (1992)'s in the
 second period,
 i.e. $\beta_{i2}>\bar{\beta_2}$, and $\beta_{i2}>\tilde{\beta_2}$. Not
 surprisingly, the depth of the market is
 bigger  than that of Hudart, Hughes
 and Levine (2001)'s and Holden and Subrahmanyam (1992)'s, i.e.
 ${1}/{\lambda_2}>{1}/{\bar{\lambda}_2}$, ${1}/{\lambda_2}>{1}/{\tilde{\lambda}_2}$.
So,  we  conclude that,  ``competition among insiders''  along with
 ``public disclosure'' have a great influence on the behavior of each
insider and the market structure.

\section{A sequential auction equilibrium}
\quad \quad
 In this section we generalize the model of two-period
trading by examining a model in which a number of trading rounds
with public disclosure taking place sequentially. The resulting
dynamic model is structured so that the equilibrium price at each
auction reflects the information contained in the past and current
order flow. Moreover, all  insiders  maximize their expected profits
in the equilibrium, taking into account their effect on price in
both the current auction and in the future auction.

 A security is traded in
$N$ sequential and its value at the end of trading is denoted by $v$
which is assumed to be normally distributed with mean prior $p_0$
and prior variance $\Sigma_0$. Let $M$ denote the number of
insiders, who are indexed by $i=1,\cdots,M.$ Each insider observes
the liquidation value in advance. We assume that the quantity traded
by noise traders at the $n$th auction is $\mu_n$, which is normally
distributed with mean $0$ and variance $\sigma_{\mu}^2$, and
$\mu_1,\cdots, \mu_N, v$ are mutually independent.  At the $n$th
auction, trading is structure in two steps as follows: In step one,
insiders and noise traders choose their trader quantity. When doing
so, every insider observes $v$ but not $u_n$. In step two,  market
makers determine the price $p_n$ to clear the market, based on their
information. Since, every insider is required to publicly disclose
his  stock trades after the $n$th auction realized, he choose the
mixed trading strategy $x_{in}=\beta_{in}(v-p_{n-1}^*)+z_n$ to
dissimulate his  private information, where $z_n$ is a sequence of
independent random variables, normally distributed with mean $0$ and
variance $\sigma_{z_n}^2$ (note that $\sigma_{z_N}^2$=0).

Let $ x_{n}$  and $ x_{in}$ denote the total order by all insiders
and the individual order by the $i$th insider at the $n$th auction
respectively. And let $\pi_{in}$ denote the total expected profit of
the $i$th ionsider from positions acquired at all future auction
$n,\cdots,N$.
 Each risk neutral insider determines his optimal trading
 strategy by a process of backward induction, in order to maximize
 his expected profits given his conjectures about the trading
 strategies of the other insiders. In the rational
 expectation equilibrium, the conjectures of each identical
insider, must be correct conditional on each trader's
 information at each auction.

We now state a proposition which provides the difference equation
 system characterizing our equilibrium.

\begin{proposition}\label{pro5}In the economy with $M$ ($M\geq 1$) insiders , there exists a unique sub-game perfect equilibrium and
this equilibrium is a sequential equilibrium.  In this equilibrium
there are constants $\beta_{in},\lambda_n,\alpha_n,\delta_n$ and
$\Sigma_n$, characterized by the following:
\begin{equation}\label{eq4.1}x_{in}=\beta_{in}(v-p_{n-1}^*)+z_{n},
\end{equation}
\begin{equation}\label{eq4.2}p_n=p_{n-1}^*+\lambda_ny_n,
\end{equation}
\begin{equation}\label{eq4.3}p_n^*=p_{n-1}^*+\gamma_nx_n,
\end{equation}
\begin{equation}\label{eq4.4}\Sigma_n=var(v|y_1,\cdots,y_n),
\end{equation}
\begin{equation}\label{eq4.5}E\{\pi_{in}|p_1^*,\cdots,p_{n-1}^*,p_1,
\cdots,p_{n-1},v\}=\alpha_{n-1}(v-p_{n-1}^*)^2+\delta_{n-1},
\end{equation}
for all auctions $n=1,\cdots,N$ and for all informed traders
$i=1,\cdots, M$.

 Given $p_0^*=p_0$, $\Sigma_0$, the constants $\beta_{in},\lambda_n,\alpha_n,\delta_n$ and
$\Sigma_n$ are the unique solution to the following difference
equation system
\begin{equation}\label{eq4.6}\alpha_{n-1}=\beta_{in}(1-\lambda_nM\beta_{in})+\alpha_n(1-\gamma_nM\beta_{in})^2,
\end{equation}
\begin{equation}\label{eq4.7}\delta_{n-1}=\frac{M-1}{2(M+1)}M\gamma_n\sigma_{z_n}^2+\delta_n,
\end{equation}
\begin{equation}\label{eq4.8}\beta_{in}=\frac{(M+1)\lambda_n\sigma_{\mu}^2}{M\Sigma_{n-1}},
\end{equation}
\begin{equation}\label{eq4.9}\lambda_n=\frac{M}{2(M+1)\alpha_n},
\end{equation}
\begin{equation}\label{eq4.10}\Sigma_n=\Sigma_{n-1}-\frac{(M+1)^2}{M}\lambda_n^2\sigma_{\mu}^2,
\end{equation}
\begin{equation}\label{eq4.11}\gamma_n=\frac{M+1}{M}\lambda_n,
\end{equation}
\begin{equation}\label{eq4.12}\sigma_{z_n}^2=\sigma_{\mu}^2-
\frac{\lambda_n^2\sigma_{\mu}^4(M+1)^2}{M\Sigma_{n-1}},
\end{equation}
for $n=1,\cdots,N-1$.  When $n=N$, we have
\begin{equation}\label{eq4.13}\delta_{N-1}=0,~~~~\alpha_{N-1}=\frac{1}{(M+1)^2\lambda_N},~~~~
\beta_{iN}=\frac{1}{(M+1)\lambda_N},~~~~\lambda_{N}=\frac{\sqrt{M\Sigma_{N-1}}}{(M+1)\sigma_{\mu}},
\end{equation}
and $\delta_N=\alpha_N=\Sigma_N=\sigma_{z_N}^2=0.$
\end{proposition}
\begin{proof} See the Appendix.
\end{proof}

{\bf Analysis of the equilibrium}: In order to compare the
imperfectly competitive case to the monopolist case,  and the public
disclosure case to without public disclosure case, we give a series
of numerical simulations using  Proposition $\ref{pro5}$ in our
paper and Proposition $2$ in Holden and Subrahmanyam (1992).

We first compare the interesting parameters $\Sigma_n$ and
$\lambda_n$ in our model with those of Hudart, Hughes and Levine
(2001)'s model. As in Kyle (1985), the parameters $\Sigma_n$ and
$\lambda_n$ are inverse measures of price efficiency and market
depth, respectively.

Figure $1$ plots the dynamic behavior of the liquidity parameter
$\lambda_n$ by holding constants $\Sigma_0=1$, $\sigma_{\mu}^2=1$,
$N=10$ fixed and varying the number of insiders
 for the cases of $M=1$, $M=2$, $M=10$ or $M=50$. It indicates that when $M=1$, i.e.,  there is a single
insider trader in the market, the liquidity parameter $\lambda_n$ is
constant, which is just the result of Hudart, Hughes and Levine
(2001)'s model. When the number of insiders is more than two, the
liquidity parameter $\lambda_n$ declines nearly to zero very rapidly
through time. As the number of insiders increases, the speed with
which they drop increases dramatically. Note also in Figure $1$ the
small $\lambda_n$ in the initial periods when the number of insiders
is large. The adverse selection (measured by $\lambda_n$) is high in
the early periods since the information content of the order flow is
high,  and negligible in the later periods because  market makers
have very little to fear from traders that have already exploited
most of their informational advantage. As the number of insiders is
increased, trading volume becomes less  at earlier auctions because
of ``competition among insiders'' and hence there is smaller adverse
selection at these auctions. This pattern is consistent with the
speed with which information is revealed,  measured by $\Sigma_n$,
which is plotted in Figure $2$.

\begin{figure}[htbp]
\begin{minipage}[c]{1.0\textwidth}
\centering
\includegraphics[width=4.0in]{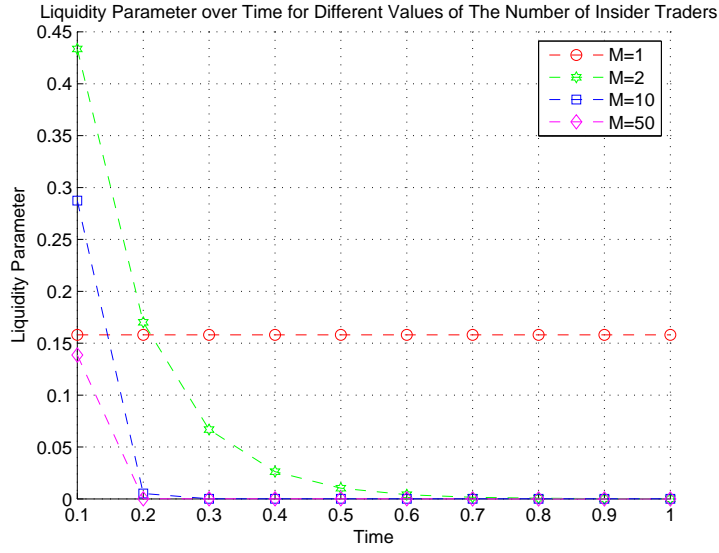}
\caption{ \small Liquidity parameter $\lambda_n$ over time for
different values of $M$, the number of informed traders. The number
of auctions is fixed at $N=10$ and the liquidity parameter at each
auction is plotted for different value of $M$, $M=1$, $M=2$, $M=10$
or $M=50$.}
\end{minipage}
\end{figure}

Figure $2$ plots the dynamic behavior of the error variance of price
$\Sigma_n$ by holding constants $\Sigma_0=1$, $\sigma_{\mu}^2=1$,
$N=20$ fixed and varying the number of insiders
 for the cases of $M=1$, $M=2$, $M=3$ or $M=4$.
It indicates that  $\Sigma_n$ declines at a linear rate when $M=1$,
 which  is consistent
with the conclusion of Hudart, Hughes
 and Levine (2001). As the
liquidity parameter $\lambda_n$, $\Sigma_n$ also declines nearly to
zero very rapidly through time. As the number of insiders increases,
the speed with which they drop increases dramatically. In fact, when
there are $3$ insiders in the market, less than $5$ percent of the
information remains to be revealed by the $1$th auction. This is
because the imperfectly competitive informed traders cannot collude
to exploit their rents slowly. The larger of the number of insiders
is, the more fierce of the competition among them.

\begin{figure}[htbp]
\begin{minipage}[c]{1.0\textwidth}
\centering
\includegraphics[width=4.0in]{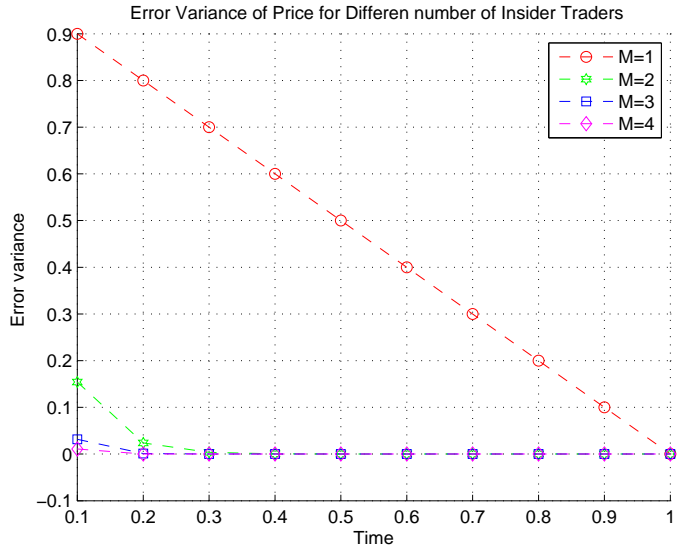}
\caption{ \small  The number of auctions is fixed at $N=10$ and the
error variance of price, $\Sigma_n$, is plotted for different values
of $M$, $M=1$, $M=2$, $M=3$ or $M=4$. }
\end{minipage}
\end{figure}

\begin{figure}[htbp]
\begin{minipage}[c]{1.0\textwidth}
\centering
\includegraphics[width=4.0in]{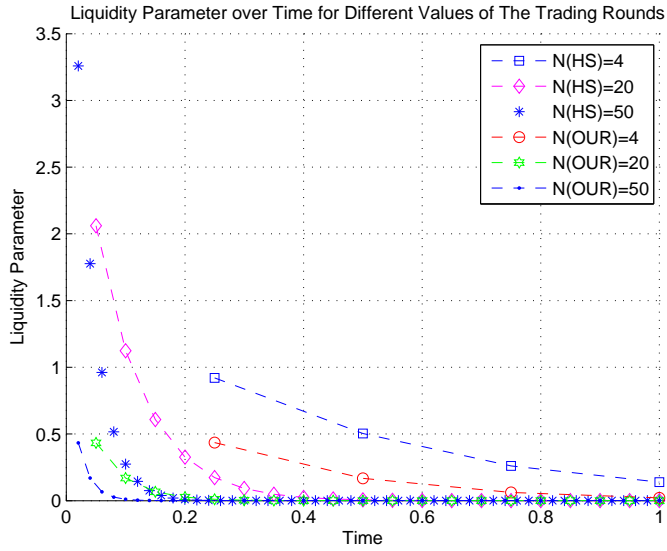}
\caption{ \small  This figure contrasts the liquidity parameter
$\lambda_n$, (i) when the insider must disclose each trade ex-post
and (ii) when no such disclosures are made. The number of insiders
 is fixed at $M=2$ and the liquidity parameter at each
auction is plotted for different value of $N$, $N=4$, $N=20$,
$N=50$. $N(HS)$ in the figure means the $N$ in Holden and
Subrahmanyam(1992) and $N(our)$ means the $N$ in our model. }
\end{minipage}
\end{figure}
\begin{figure}[htbp]
\begin{minipage}[c]{1.0\textwidth}
\centering
\includegraphics[width=4.0in]{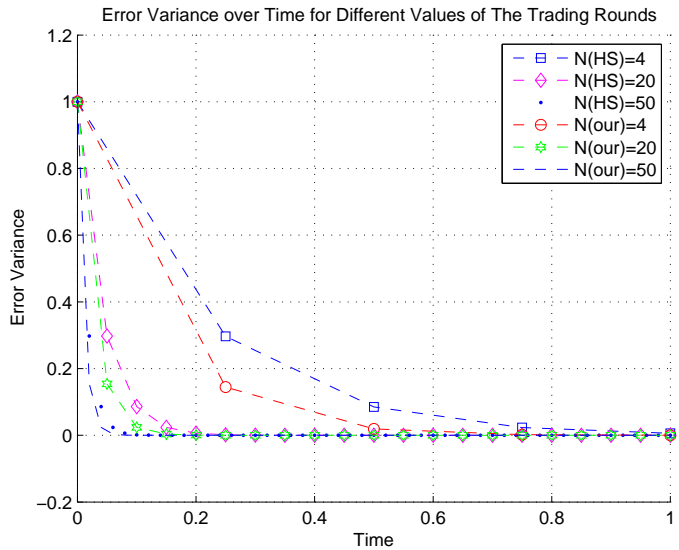}
\caption{ \small This figure contrasts the error variance of price,
$\Sigma_n$,  in our model and in Holden and Subrahmanyam (1992)'s
model, over time for different values of $N$, the number of
 trading rounds. The number of insiders  is fixed at $M=2$ and the
error variance of price at each auction is plotted for different
value of $N$, $N=4$, $N=20$, $N=50$. $N(HS)$ in the figure means the
$N$ in Holden and Subrahmanyam (1992) and $N(our)$ means the $N$ in
our model.}
\end{minipage}
\end{figure}
\begin{figure}[htbp]
\begin{minipage}[c]{1.0\textwidth}
\centering
\includegraphics[width=4.0in]{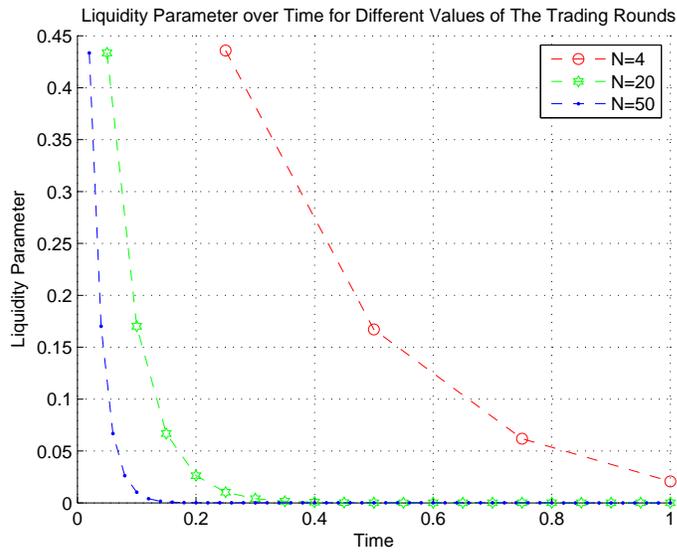}
\caption{ \small  Liquidity parameter $\lambda_n$ over time for
different values of $N$, the number of
 trading rounds. The number of insiders  is fixed at $M=2$ and the
liquidity parameter at each auction is plotted for different value
of $N$, $N=4$, $N=20$, $N=40$.}
\end{minipage}
\end{figure}
\begin{figure}[htbp]
\begin{minipage}[c]{1.0\textwidth}
\centering
\includegraphics[width=4.0in]{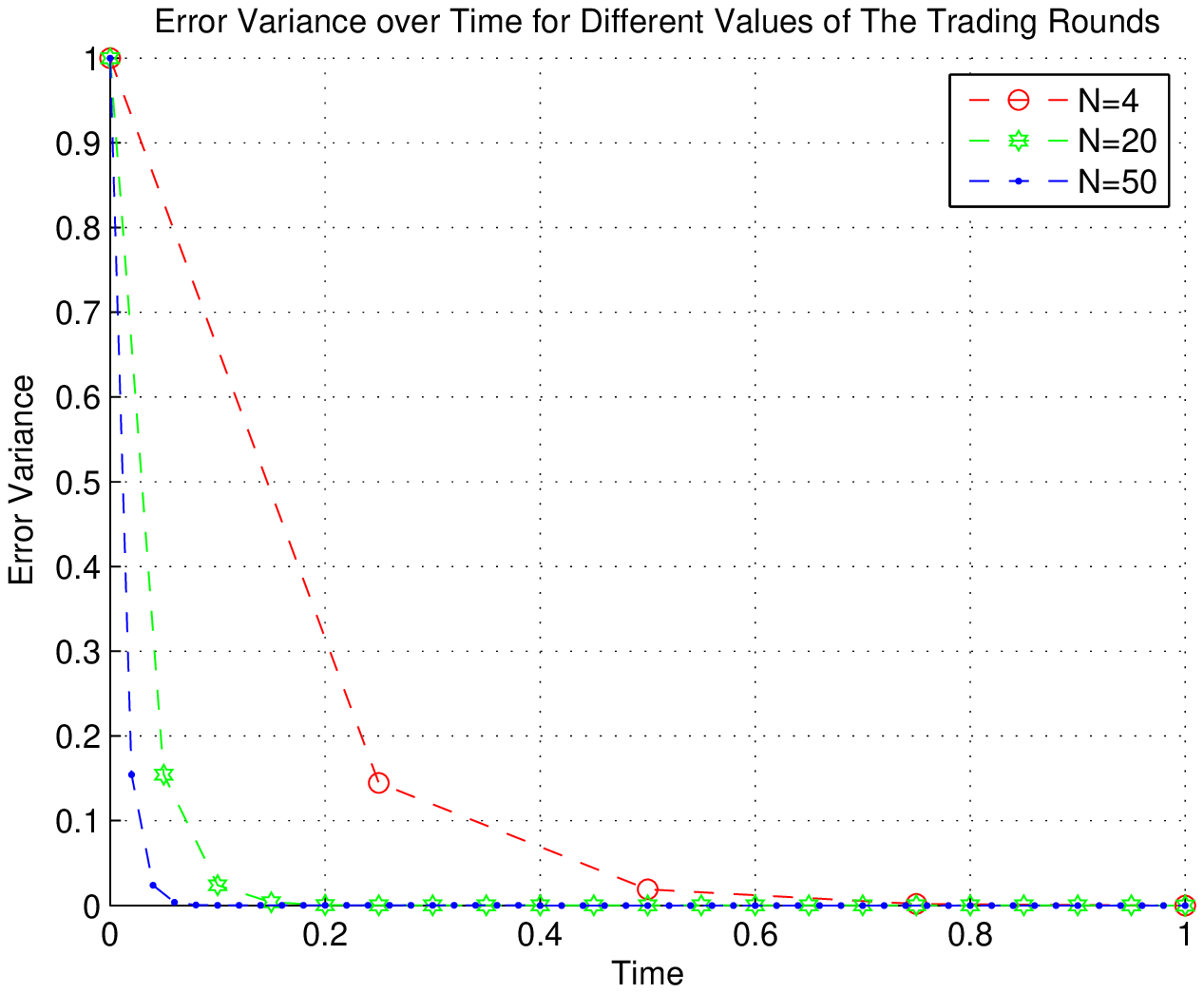}
\caption{ \small   The number of insiders is fixed at $M=2$ and the
error variance of price, $\Sigma_n$, is plotted for different values
of $N$, the number of auctions.}
\end{minipage}
\end{figure}

Figure $3$ and $4$ plot $\lambda_n$ and $\Sigma_n$ in our model and
in Holden and Subrahmanyam (1992)'s, respectively, by holding
constants $\Sigma_0=1$, $\sigma_{\mu}^2=1$, $M=2$ fixed and varying
 numbers of auctions, $N=4$,
$N=20$, or $N=50.$ In order to see  the behavior of $\lambda_n$ and
$\Sigma_n$ in our model clearly,  Figure $5$ and Figure $6$ again
plot the liquidity parameter and the error variance for varying
numbers of trading rounds, $N=4$, $N=20$, $N=50$, respectively.

Figure $3$ and Figure $4$ illustrate the contrast in the equilibria
(i) when the insider must disclose each trade ex-post, and (ii) when
no such disclosure are made. Figure $3$ illustrates that when the
number of trading rounds fixed, the liquidity parameter $\lambda_n$
with public disclosure of insider trades is smaller than that
without public disclosure. That is to say the adverse selection is
low because of ``public disclosure''.  Also, as the number of
trading rounds per unit time becomes larger, the $\lambda_n$ of
Holden and Subrahmanyam (1992) in the initial periods become larger,
but it is not the case in our model, which is potted clearly in
Figure $5$. In  Holden and Subrahmanyam (1992)  the adverse
selection (measured by $\lambda_n$) is high in the earlier periods
because the information content of the order flow is high, and
negligible in the later periods because market makers have very
little to fear from traders that have already exploited most of
their informational advantage. As the number of auctions is
increased, trading becomes more concentrated at earlier auctions and
hence there is greater adverse selection at these auctions. But
under the requirement of ``public disclosure'', market makers know
that insiders may disclose their trading and may choose a mixed
strategy to dissimulate their information,  the adverse selection is
the same at the initial auction even if the number of auctions
increases.

Figure $4$ illustrates that as the number of trading rounds per unit
time becomes very large, $\Sigma_n$ declines nearly to zero very
rapidly both in the case of ``public disclosure'' and the case
``without public disclosure''. And as the number of auctions
increases, the speed with which they drop increases dramatically.
But when the number of trading rounds fixed, the drop speed is
faster under the requirement of ``public disclosure''. That is to
say, the
 ``public disclosure'' and  aggressive competition among insiders lead to most of the
private information to be incorporated into the price in the early
auctions.

\section{A convergence result--Near continuous trading}
\quad \quad In this section,  we will study  the properties of the
discrete model of sequential trading when the interval between
auctions is vanishingly small, and analyze  the behavior of the
equilibrium when the market approaches continuous trading. Our main
results, given in Theorem $\ref{th5}$, concern the behavior of the
equilibrium when the market approaches continuous trading. Unlike
Back (1992) and Back, Cao, and Willard (2000) e.t.c  formulating the
continuous time model directly, we use the  discrete models to
approach continuous model, that is, use the discrete results
Proposition ${\ref{pro5}}$ and then take the limit.

 We assume there is $N$ auctions
in the interval $[0,1]$, and the interval between auctions is the
same.

In order to get the continuous result we first give  the following
key proposition, the proof of which is given in the appendix.
\footnote{ The method that we used here to get the convergence
result is different from the method used by Holden and Subrahmanyam
(1992) and Kyle (1985).  }
\begin{proposition}\label{pro6}For any fixed $N$, the sequence $\{a_n\}$
described by the equation $(\ref{eq5.41})$ satisfy
\begin{equation}{\label{eq6.1}}\frac{a_{n-1}^2}{a_n^2}\leq 1
\end{equation}
where $1\leq n\leq N$.
\end{proposition}

The following theory is one of our main results:
\begin{theorem}\label{th5}
As the interval between auctions in the discrete model becomes
uniformly small, (i.e.$N\rightarrow +\infty$), the parameters
characterized by the unique sequential auction equilibrium in
Proposition
 $\ref{pro5}$ is convergent.  More precisely,
for any $t\in (0,1)$ \footnote{ Because we assume that every inside
trader must disclose his trading volume after the fact but not
verification afterward by others, every insider uses his information
completely after the trade finished, i.e. $\Sigma_1=0, \lambda_1=0$.
}, choose $N_1<N_2<\cdots<N_j<\cdots\rightarrow
  +\infty$, and satisfy   $2+(1-t)N_j\leq (1-t)N_{j+1}<N_j-1, j=1,2,\cdots.$
For every  $N_j$, let
 $n(N_j)=[tN_j] $,  $j=1, \cdots, $  where $[x]$ denote the integer
 part of
 $x$. When
   $j\rightarrow +\infty$,  we have $\Sigma_{n(N_j)}^{(N_j)} \rightarrow
\Sigma_t$, $\lambda_{n(N_j)}^{(N_j)} \rightarrow \lambda_t$,
$\beta_{in(N_j)}^{(N_j)} \rightarrow
\beta_{it}$,${\sigma_{z_{n(N_j)}}^{(N_j)}}^2 \rightarrow
\sigma_{z_t}^2$
 and

 (i)When $M=1$,
$\Sigma_t$, $\lambda_t$, $\beta_{it}$ , $\sigma_{z_t}^2$ satisfy
$$\Sigma_t=(1-t)\Sigma_0,~~~~\lambda_t=0,$$
$$\beta_{it}=0~~(i=1),~~~~~\sigma_{z_t}^2=\sigma_{\mu}^2.$$

(ii) When $M\geq 2$, $\Sigma_{n(N_j)}^{(N_j)}$ satisfies
\begin{equation}\label{eq**}\exp\left\{[tN_j]\log\left(1-\frac{A}{M}\right)+o(N_j)\right\}
\leq\frac{\Sigma_{n(N_j)}^{(N_j)}}{\Sigma_0}\leq
\exp\left\{[tN_j]\log\left(1-\frac{A}{M}\right)\right\}
\end{equation}
i.e. $\frac{\Sigma_{n(N_j)}^{N_j}}{\Sigma_0}\approx\exp
\left\{[tN_j]\log\left(1-\frac{A}{M}\right)\right\}$, and
\begin{equation}\label{eq*}\lim_{N_j\rightarrow+\infty}\lambda_1^{(N_j)}=\frac{\sqrt{\Sigma_0}A}{(M+1)\sigma_{\mu}},
\end{equation}
\begin{equation}\label{eq***}\lim_{N_j\rightarrow+\infty}\beta_1^{(N_j)}=\frac{A\sigma_{\mu}}{M\sqrt{\Sigma_0}},
\end{equation}
furthermore, $\Sigma_t$, $\lambda_t$, $\beta_{it}$ ,
$\sigma_{z_t}^2$ satisfy
$$\Sigma_t=0 , ~~~~~\lambda_t=0,$$
$$\beta_{it}=+\infty,~~~\sigma_{z_t}^2=\frac{\sigma_{\mu}^2}{M}(1-\frac{A}{M}),$$
where $A$ is  the unique root of $f(x)\doteq
(M-1)^2x^3-4(M+1)(M-1)^2x^2+(M^2-1)(6M^2-4)x+(M+1)^2(4M^2-3M^3)=0$
in $(0,M)$.
\end{theorem}
\begin{proof}See the Appendix.
\end{proof}

Theorem $\ref{th5}$ implies that, when $M=1$, the speed of insiders'
information incorporated into the price is  a linear function of the
time $t$. This is a limiting result of Hudart, Hughes and Levine
(2001).  When $M\geq 2$ , the speed of the insiders' information is
not incorporated into the price is approximate a exponential
function of the time $[tN_j]$. When $j\rightarrow +\infty$,
$\Sigma_t=0$, $\lambda_t=0$ for all $t\in (0,1),$ that is to say, as
the number of auctions goes to infinity, all information is revealed
immediately and the market depth goes to infinity immediately. These
results
 described in Theorem $\ref{th5}$ are consistent with the results
 numerical illustrated in section $2$.

But another class of limits (i.e.
$\lim_{N_j\rightarrow+\infty}\lambda_1^{(N_j)}=\frac{\sqrt{\Sigma_0}A}{(M+1)\sigma_{\mu}}$
and
$\lim_{N_j\rightarrow+\infty}\beta_1^{(N_j)}=\frac{A\sigma_{\mu}}{M\sqrt{\Sigma_0}}$)
show that, as the total number of auction goes to infinity, the
market depth at the first auction is a constant, and the trading
intensity is also a constant. However, Holden and Subrahmanyam
(1992) find that the market depth at the first auction goes to zero
and trading intensity $\beta_1$ goes to infinity. The intuition for
the contrast is that ``public disclosure'' enable  higher adverse
selection in the market.

\section{Conclusion}
\quad \quad
 In this paper,  we consider a  model in which there are
at least two informed traders in the
 market,  with the requirement that  insiders  publicly disclose their
 stock trades after the fact but not verification by others.
  In equilibrium,
the ``public disclosure'' and  ``competition among
insiders''assumptions  have great effects on each insider's trading
intensity, the market depth and the effectiveness of the price.

We have shown that  the contrast between our results and those of
 Hudart, Hughes and Levine (2001) and  the contrast between our results and
those of Holden and Subrahmanyan (1992) are far from trivial. In
particular, in our model, as the interval between auctions
approaches zero, the market approaches the perfectly competitive
outcome of full information revelation and infinite market depth
almost immediately, in contrast to the case of a monopolist
considered by Hudart, Hughes and Levine (2001), wherein market depth
is constant at all times and information is revealed at a linear
speed. Moreover, in the two-period trading, the market depth
 of the first auction in
our model is smaller
 than Hudart, Hughes
 and Levine (2001)'s when the number of insiders
 is less than $5$, and  the opposite conclusion holds if the number of insiders is more than $5$. That is to say,
 whether market makers set the marginal cost of the first period traders lower or higher depends on the number of insiders.
 If the number of insiders is big enough (bigger than $5$), the adverse selection becomes small.
Comparing with
 Holden and Subrahmanyan(1992)'s model,   each  insider plays a mixed
strategy in every round except the last one, because of the
existence of ``public disclosure'', leading  to the accelerated
price discovery  and  higher market depths.
 Furthermore,
in the sequential auction equilibrium, market depths become infinite
and all private information is revealed immediately when the time
interval between auctions approaches zero. The speed of the
revelation is faster than that of Holden and Subrahmanyan (1992),
and we give the exact speed of the revelation in Theorem \ref{th5}.

Moreover, we give the near-continuous trading results by starting
with discrete time and then taking the limit, rather than
formulating the model directly in continuous time.  We find that the
speed of insiders' information incorporated into the price is  a
linear function of the time $t$ ($t\in(0,1)$) when there is one
insider in the market. However, if there are more than two insiders,
the speed of the insiders' information is not incorporated into the
price is approximate a exponential function of the time $[tN]$,
where $t\in(0,1)$ is the time and $N$ is the total number of
auctions in $(0,1).$ That is to say, when there are more than two
insiders in the market all information is revealed immediately and
the market depth goes to infinity immediately as the number of
auctions goes to infinite where the market is strong-form efficient.
An interesting find is that  as the total number of auction goes to
infinity, the market depth and the trading intensity at the first
auction are all constants while Holden and Subrahmanyam (1992) find
that the market depth at the first auction goes to zero and the
trading intensity of the first period goes to infinity. The
intuition for the contrast is that ``public disclosure'' enable less
adverse selection in the market.

Our results have two important implications. Firstly, under the
requirement of ``public disclosure'', the market can be close to
strong-form efficiency  even when the presence of just two
noncooperative privately informed traders is assumed. Secondly,
despite being close to efficiency, the market depth at the first
auction is a constant. These implications are in sharp contrast to
previous literature.

 Our model not only shows that  insiders are able to
profit from long-lived private information even though they must
disclose their trades at the end of each trading, but also can be
applied to lots of intra-day phenomena, such as adverse selection
problem and so on.

\section{Appendix}

We first recall a well-known regression result which  will be used.
\noindent \begin{lemma}{\label{lem a}}Let $X_1$ and $X_2$ be two
normal random vectors, $
\left(\begin{array}{c}X_1\\X_2\end{array}\right)\sim N(\mu,\Sigma) $
with $\mu=\left(\begin{array}{c}\mu_1\\\mu_2\end{array}\right)$,
$\Sigma=
\left(\begin{array}{cc}\Sigma_{11}&\Sigma_{12}\\
\Sigma_{21}&\Sigma_{22}\end{array}\right)$.
 Then the random variable $X_1$ conditional on $X_2$  has a normal distribution. More precisely ,
 $$E[X_1|X_2]=\mu_1+\Sigma_{12}\Sigma_{22}^{-1}(X_2-\mu_2),~~~~
Var(X_1|X_2)=\Sigma_{11}-\Sigma_{12}\Sigma_{22}^{-1}\Sigma_{21}.
 $$
 \end{lemma}

{\bf Proof of Proposition \ref{pro1}}: Let  $N$ in Proposition $1$
of Holden and Subrahmanyam (1992) equal $2$ and $\Delta t_n$
$(n=1,\cdots,N.)$ equal $1$,
 we have the following relationships :
\begin{equation}{\label{eq5.1}}\alpha_1=\frac{1-\alpha_2\lambda_2}{\lambda_2[M(1-2\alpha_2\lambda_2)+1]^2},\end{equation}
\begin{equation}{\label{eq5.2}}\alpha_2=0,\end{equation}
\begin{equation}{\label{eq5.3}}\beta_1=\frac{1-2\alpha_1\lambda_1}{\lambda_1[M(1-2\alpha_1\lambda_1)+1]},\end{equation}
\begin{equation}{\label{eq5.4}}\beta_2=\frac{1}{\lambda_2(M+1)},\end{equation}
\begin{equation}{\label{eq5.5}}\lambda_1=\frac{M\beta_1\Sigma_1}{\sigma_{\mu}^2},\end{equation}
\begin{equation}{\label{eq5.6}}\lambda_2=\frac{1}{\sigma_{\mu}}\left(\frac{M\Sigma_2}{M+1}\right)^{\frac{1}{2}},\end{equation}
\begin{equation}{\label{eq5.7}}\Sigma_2=(1-M\beta_2\lambda_2)\Sigma_1,\end{equation}
\begin{equation}{\label{eq5.8}}\Sigma_1=(1-M\beta_1\lambda_1)\Sigma_0.\end{equation}
Let $k\doteq \frac{\lambda_1}{\lambda_2}$,
 from (\ref{eq5.1}) and
(\ref{eq5.2}) we  get
\begin{equation}{\label{eq5.9}}\alpha_1=\frac{1}{\lambda_2(M+1)^2}.
\end{equation}
Combining (\ref{eq5.3}) and (\ref{eq5.9}),  we  get the expression
of $\beta_1$
$$\beta_1=\frac{1-\frac{2k}{(M+1)^2}}{\lambda_1[M(1-\frac{2k}{(M+1)^2})+1]}=\frac{(M+1)^2-2k}{\lambda_1[(M+1)^3-2kM]}.$$
The expression of $\beta_1$ and (\ref{eq5.8}) imply
\begin{equation}{\label{eq5.10}}\Sigma_1=\left(1-\frac{M(M+1)^2-2kM}{(M+1)^3-2kM}\right)
\Sigma_0 =\frac{(M+1)^2\Sigma_0}{(M+1)^3-2kM}.\end{equation}
(\ref{eq5.4}),(\ref{eq5.7}) and (\ref{eq5.10}) give:
\begin{equation}{\label{eq5.11}}\Sigma_2=\frac{1}{M+1}\Sigma_1=\frac{(M+1)\Sigma_0}{(M+1)^3-2kM}.
\end{equation}
 (\ref{eq5.5}) and  (\ref{eq5.10}) imply:
\begin{equation}{\label{eq5.12}} \lambda_1^2
=
\frac{M(M+1)^2[(M+1)^2-2k]\Sigma_0}{[(M+1)^3-2kM]^2\sigma_{\mu}^2}.
\end{equation}
Combining (\ref{eq5.6}) and (\ref{eq5.11})
\begin{equation}{\label{eq5.13}}
\lambda_2^2=\frac{1}{\sigma_{\mu}^2}\frac{M\Sigma_0}{(M+1)^3-2kM.
}\end{equation} From the above two equations it is easy to get
\begin{equation}k^2\doteq
\frac{\lambda_1^2}{\lambda_2^2}=\frac{(M+1)^2[(M+1)^2-2k]}{(M+1)^3-2kM}.
\end{equation}
The polynomial $2Mk^3-(M+1)^3k^2-2(M+1)^2k+(M+1)^4=0$ has three
roots, but from the second order condition of Proposition $1$ in
Holden and Subrahmanyam (1992), $\lambda_1>0$, $\lambda_2>0$. It is
easy to show that the polynomial only has one root satisfying
conditions $\lambda_1>0$ and $\lambda_2>0$.

{\bf Proof of Proposition \ref{pro3}}:\ \ \ From  Proposition $2$ of
Hudart, Hughes and Levine (2001), we get
$$\bar{\lambda}_1=\bar{\lambda}_2=\frac{1}{2\sigma_{\mu}}\sqrt{\frac{\Sigma_0}{2}},$$
$$\bar{\beta}_1=\frac{\sigma_{\mu}}{\sqrt{2\Sigma_0}},  ~~~\bar{\beta}_2=\frac{\sqrt{2}\sigma_{\mu}}{\sqrt{\Sigma_0}}$$
$$\bar{\Sigma}_1=\frac{1}{2}\Sigma_0.$$
So ,
$$\frac{\lambda_1}{\bar{\lambda}_1}=\sqrt{\frac{8M^3}{4+M^2(M+1)^2}}=\left\{\begin{array}{c}  >1,\ \ \  2 \leq M \leq
5\\
<1, \ \ \ M>5\end{array}\right.,$$
$$\frac{\lambda_2}{\bar{\lambda}_2}=\frac{4}{(M+1)^2}\sqrt{\frac{2M}{[4+M^2(M+1)^2]}}<1,$$
$$\frac{\beta_{i1}}{\bar{\beta}_1}=(M+1)\sqrt{\frac{2M}{4+M^2(M+1)^2}}<1,$$
$$\frac{\beta_{i2}}{\bar{\beta}_2}=\sqrt{\frac{4+M^2(M+1)^2}{8M}}>1,$$
$$\frac{\Sigma_1}{\bar{\Sigma}_1}=\frac{8}{4+M^2(M+1)^2}<1.$$

{\bf Proof of Proposition \ref{pro4}}:  Let the $M$ in Proposition
\ref{pro1}  equal   $2$,  we have
$$\tilde{\lambda}_1=\frac{\sqrt{2\Sigma_0(1-\frac{2k}{9})}}{\sigma_{\mu}(3-\frac{4k}{9})},
   ~~~\tilde{\lambda}_2=\frac{1}{3\sigma_{\mu}}\sqrt{\frac{2\Sigma_0}{3-\frac{4k}{9}}},$$
$$\tilde{\beta}_1=\frac{1-\frac{2k}{9}}{\tilde{\lambda}_1(3-\frac{4k}{9})}=
\sigma_{\mu}\sqrt{\frac{1-\frac{2k}{9}}{2\Sigma_0}},~~~\tilde{\beta}_2=\frac{1}{3\tilde{\lambda}_2}=
\sigma_{\mu}\sqrt{\frac{3-\frac{4k}{9}}{2\Sigma_0}},$$
$$\tilde{\Sigma}_1=\frac{\Sigma_0}{3-\frac{4k}{9}},$$
and the $k\doteq\frac{\lambda_1}{\lambda_2}$ satisfies:
$4k^3-27k^2-18k+81=0$, so $k=6.9790,-1.8218$, or $1.5927.$  By the
second order condition we have $\lambda_1>0$, so $k$ must satisfy
$0<k<4.5$.  So $k=1.5927$.
 Let the $M$ in Proposition \ref{pro2} equal $2$, we have
 $$\lambda_1=\frac{\sqrt{\Sigma_0}}{3\sigma_{\mu}\sqrt{5}},~~~ \lambda_2=\frac{\sqrt{\Sigma_0}}{3\sigma_{\mu}\sqrt{5}},$$
 $$\beta_{i1}=\frac{3\sigma_{\mu}\sqrt{5}}{10\sqrt{\Sigma_0}},~~~ \beta_{i2}=\frac{\sigma_{\mu}\sqrt{5}}{\sqrt{\Sigma_0}},$$
$$\Sigma_1=\frac{1}{10}\Sigma_0.$$
So, it is easy to get
$$\frac{\lambda_1}{\tilde{\lambda}_1}=\frac{\sqrt{\Sigma_0}}{\sigma_{\mu}\sqrt{5}}\times
\frac{\sigma_{\mu}(3-\frac{4k}{9})}{\sqrt{2\Sigma_0(1-\frac{2k}{9})}}\approx\frac{2.39}{\sqrt{6.5}}<1,$$
$$\frac{\lambda_2}{\tilde{\lambda}_2}=\frac{\sqrt{\Sigma_0}}{3\sigma_{\mu}\sqrt{5}}\times
\frac{3\sigma_{\mu}(3-\frac{4k}{9})}{\sqrt{2\Sigma_0}}\approx\frac{\sqrt{2.3}}{\sqrt{10}}<1,$$
$$\frac{\beta_{i1}}{\tilde{\beta}_1}=\frac{3\sigma_{\mu}\sqrt{5}}{10\sqrt{\Sigma_0}}\times
\frac{\sqrt{2\Sigma_0}}{\sigma_{\mu}\sqrt{1-\frac{2k}{9}}}\approx
\frac{3}{\sqrt{7.4}}>1,$$
$$\frac{\beta_{i2}}{\tilde{\beta}_2}=\frac{\sigma_{\mu}\sqrt{5}}{\sqrt{\Sigma_0}}\times
\frac{\sqrt{2\Sigma_0}}{\sigma_{\mu}\sqrt{3-\frac{4k}{9}}}\approx
\frac{\sqrt{10}}{\sqrt{2.3}}>1.$$

{\bf {Proof of Proposition \ref{pro5}}}: Applying the principal of
backward induction, we can write
 the insider $i$'s last period ($N$th period) optimization problem
 for given $x_{i1},\cdots,x_{i,N-1},p_1^*,\cdots,p_{N-1}^*$ as $x_{iN} \in
  arg\max_{x}E[x(v-p_N)|x_{j1},\cdots,x_{j,N-1},1\leq j \leq M , p_1^*,\cdots,p_{N-1}^*,v]$,
  where
$$\begin{aligned}
&E[x(v-p_N)|x_{j1},\cdots,x_{j,N-1},1\leq j \leq M
,p_1^*,\cdots,p_{N-1}^*,v]\\=&E[x(v-p_{N-1}^*-\lambda_Nx-\lambda_N
(M-1)\bar{x}_{iN})|x_{j1},\cdots,x_{j,N-1},1\leq j \leq M ,p_1^*,\cdots,p_{N-1}^*,v]\\
=&x[v-p_{N-1}^*-\lambda_Nx-\lambda_N (M-1)\bar{x}_{iN}],
\end{aligned}
$$
in the above  $\bar{x}_{iN}$ represents the particular informed
trader's conjecture of the average of the other informed traders'
strategies, i.e. $\bar{x}_{iN}=\frac{1}{M-1}E[\sum_{j\neq
i}x_{jN}|x_{j1},\cdots,x_{j,N-1},1\leq j \leq M
,p_1^*,\cdots,p_{N-1}^*,v]$. By the first order condition we can
have
$$v-p_{N-1}^*-2\lambda_Nx_{iN}-\lambda_N (M-1)E[\sum_{j\neq
i}x_{jN}|x_{j1},\cdots,x_{j,N-1},1\leq j \leq M
,p_1^*,\cdots,p_{N-1}^*,v]=0.$$ By the theory of solving linear
equations, we know $x_{1N}=x_{2N}=\cdots=x_{MN}$,
$\bar{x}_{iN}=x_{iN}$, and
$x_{iN}=\frac{1}{(M+1)\lambda_N}(v-p_{N-1}^*)$. So,
$$\beta_{iN}=\frac{1}{(M+1)\lambda_N},$$
and
$$\begin{aligned}
&E(\pi_{iN}|x_{i1},\cdots,x_{i,N-1},p_1^*,\cdots,p_{N-1}^*,v)\\=&\max_xE[x(v-p_N)|x_{i1},
\cdots,x_{i,N-1},p_1^*,\cdots,p_{N-1}^*,v]\\=&x_{iN}(v-p_{N-1}^*-\lambda_NMx_{iN})\\=&
\frac{1}{(M+1)\lambda_N}(v-p_{N-1}^*)[v-p_{N-1}^*-\frac{M}{(M+1)}(v-p_{N-1}^*)]\\=&
\frac{1}{(M+1)^2\lambda_N}(v-p_{N-1}^*)^2,
\end{aligned}
$$
which  implies
 $\alpha_{N-1}=\frac{1}{(M+1)^2\lambda_N},\delta_{N-1}=0$.
 By the market efficient condition we can have $\lambda_N=\frac{\beta_{iN}
 \Sigma_{N-1}}{\beta_{iN}^2\Sigma_{N-1}+\sigma_{\mu}^2}$, i.e.,
 $\lambda_N=\frac{\sqrt{M\Sigma_{N-1}}}{(M+1)\sigma_{\mu}}$.

Now make the inductive hypothesis that for constants $\alpha_n$ and
$\delta_n$
\begin{equation}{\label{eq5.14}}
E(\pi_{i,n+1}|p_1^*,\cdots,p_{n}^*,p_1,\cdots,p_{n},v)=\alpha_{n}(v-p_{n}^*)^2+\delta_{n}.
\end{equation}
Since $\pi_{in}$ is given recursively by
$\pi_{in}=(v-p_n)x_{in}+\pi_{i,n+1}$, we obtain
\begin{equation}{\label{eq5.15}}
\begin{aligned}&E(\pi_{i,n}|p_1^*,\cdots,p_{n-1}^*,p_1,\cdots,p_{n-1},v)\\
=&\max_{x}E\{(v-p_n)x+\alpha_n(v-p_n^*)^2+\delta_n|p_1^*,\cdots,p_{n-1}^*,p_1,\cdots,p_{n-1},v\}.
\end{aligned}
\end{equation}
In a linear equilibrium, $p_n$ and $p_n^*$ are given by
\begin{equation}{\label{eq5.16}}p_n=p_{n-1}^*+\lambda_ny_n,
\end{equation}
\begin{equation}\label{eq5.17}p_n^*=p_{n-1}^*+\gamma_nx_n.
\end{equation}

Now, $x_n$ can be written as $x_{in}+(M-1)\bar{x}_{in}$, where
$\bar{x}_{in}$ represents the particular informed trader's
conjecture of the average of the other informed traders' strategies,
so we have
\begin{equation}{\label{eq5.18}}
\begin{aligned}&E(\pi_{i,n}|p_1^*,\cdots,p_{n-1}^*,p_1,\cdots,p_{n-1},v)\\
=&\max_{x}E\{(v-p_{n-1}^*-\lambda_ny_n)x+\alpha_n(v-p_{n-1}^*-\gamma_nx_n)^2
+\delta_n|p_1^*,\cdots,p_{n-1}^*,p_1,\cdots,p_{n-1},v\}\\
=&\max_{x}\{(v-p_{n-1}^*-\lambda_nx-\lambda_n(M-1)\bar{x}_{in})x+\alpha_n(v-p_{n-1}^*-\gamma_nx
-\gamma_n(M-1)\bar{x}_{in})^2 \\& +\delta_n\}.
\end{aligned}
\end{equation}
 By the first order condition, we can get
\begin{equation}{\label{eq5.19}}v-p_{n-1}^*-2\lambda_nx_{in}-\lambda_n(M-1)\bar{x}_{in}-2\gamma_n\alpha_n[v-p_{n-1}^*
-\gamma_nx_{in} -\gamma_n(M-1)\bar{x}_{in}]=0,
\end{equation}
and  the second order condition is
\begin{equation}{\label{eq5.20}}\lambda_n\geq
\frac{2\gamma_n^2\alpha_nM}{(M+1)}.
\end{equation}
Using  the strategy symmetry for each insider again, we have
$\bar{x}_{in}=x_{in}$. $(\ref{eq5.19})$ can be rewritten as
\begin{equation}{\label{eq5.21}}(1-2\gamma_n\alpha_n)(v-p_{n-1}^*)+[2\gamma_n^2\alpha_nM-(M+1)\lambda_n]x_{in}=0.
\end{equation}
Since the insider must be willing to play a mixed strategy,
$x_{in}=\beta_{in}(v-p_{n-1}^*)+z_n$, and $z_n$ is independent with
$v-p_{n-1}^*$, we have the following two equations by substituting
the expression of $x_{in}$ into $(\ref{eq5.21})$,
\begin{equation}{\label{eq5.22}}1-2\gamma_n\alpha_n=0,
\end{equation}
\begin{equation}{\label{eq5.23}}2\gamma_n^2\alpha_nM-(M+1)\lambda_n=0.
\end{equation}
Using  $x_{in}=\beta_{in}(v-p_{n-1}^*)+z_n$ again, we  get
\begin{equation}{\label{eq5.24}}
\begin{aligned}&E(\pi_{i,n}|p_1^*,\cdots,p_{n-1}^*,p_1,\cdots,p_{n-1},v)\\
=&E\{[v-p_{n-1}^*-\lambda_nM\beta_{in}(v-p_{n-1}^*)-\lambda_nMz_n-\lambda_n\mu_n][\beta_{in}(v-p_{n-1}^*)+z_n]\\
&+\alpha_n[v-p_{n-1}^*-\gamma_nM\beta_{in}(v-p_{n-1}^*)-\gamma_nMz_n]^2
+\delta_n|p_1^*,\cdots,p_{n-1}^*,p_1,\cdots,p_{n-1},v\}\\
=&[\beta_{in}(1-\lambda_nM\beta_{in})+\alpha_n(1-\gamma_nM\beta_{in})^2](v-p_{n-1}^*)^2+
\alpha_n\gamma_n^2M^2\sigma_{z_n}^2-\lambda_nM\sigma_{z_n}^2+\delta_n,
\end{aligned}
\end{equation}
so,
\begin{equation}{\label{eq5.25}}\alpha_{n-1}=\beta_{in}(1-\lambda_nM\beta_{in})+\alpha_n(1-\gamma_nM\beta_{in})^2,
\end{equation}
$$\delta_{n-1}=\alpha_n\gamma_n^2M^2\sigma_{z_n}^2-\lambda_nM\sigma_{z_n}^2+\delta_n.
$$
By (\ref{eq5.22}) and (\ref{eq5.23}) we  get
\begin{equation}{\label{eq5.26}}
\begin{aligned}\delta_{n-1}
=&\frac{M-1}{2(M+1)}M^2\gamma_n\sigma_{z_n^2}+\delta_n,
\end{aligned}
\end{equation}
(\ref{eq5.25}) and (\ref{eq5.26}) are just (\ref{eq4.6}) and
(\ref{eq4.7}).

 Now by the market efficient condition we  get
\begin{equation}{\label{eq5.27}}
\begin{aligned}p_n-p_{n-1}^*
=&E(v-p_{n-1}^*|y_1,\cdots,y_n,p_1^*,\cdots,p_{n-1}^*,p_1,\cdots,p_{n-1})\\
=&E[v-p_{n-1}^*|M\beta_{in}(v-p_{n-1}^*)+Mz_n+\mu_n].
\end{aligned}
\end{equation}
By Lemma {\ref{lem a}} we have
\begin{equation}{\label{eq5.28}}\lambda_n=\frac{M\beta_{in}\Sigma_{n-1}}{M^2\beta_{in}^2\Sigma_{n-1}
+M^2\sigma_{z_n}^2+\sigma_{\mu}^2}.
\end{equation}
Using the same method, we  get the expression of $\gamma_n$ as
\begin{equation}{\label{eq5.29}}\gamma_n=\frac{M\beta_{in}\Sigma_{n-1}}{M^2\beta_{in}^2\Sigma_{n-1}
+M^2\sigma_{z_n}^2}.
\end{equation}
From  the equation (\ref{eq5.22}) and (\ref{eq5.23}) it is easy to
get that $\lambda_n=\frac{M\gamma_n}{M+1}$. Combining the equation
(\ref{eq5.28}) and (\ref{eq5.29}), we have
\begin{equation}{\label{eq5.30}}M^2\beta_{in}^2\Sigma_{n-1}
+M^2\sigma_{z_n}^2=M\sigma_{\mu}^2.
\end{equation}

Now we consider $\Sigma_n$, the  measure of the informativeness of
price,
\begin{equation}{\label{eq5.31}}
\begin{aligned}\Sigma_n=&Var(v|x_1,\cdots,x_n,p_1,\cdots,p_{n-1},p_1^*,\cdots,p_{n-1}^*)\\
=&Var(v-p_{n-1}^*|M\beta_{in}(v-p_{n-1}^*)+Mz_n),
\end{aligned}
\end{equation}
using Lemma {\ref{lem a}} again, we have
\begin{equation}{\label{eq5.32}}
\begin{aligned}\Sigma_n=&\Sigma_{n-1}-\frac{M^2\beta_{in}^2\Sigma_{n-1}^2}{M^2\beta_{in}^2
\Sigma_{n-1}+M^2\sigma_{z_n}^2}\\
=&\Sigma_{n-1}-\frac{(M+1)^2}{M}\lambda_n^2\sigma_{\mu}^2.
\end{aligned}
\end{equation}
The equation
$\alpha_n=\frac{1}{2\gamma_n}=\frac{M}{2(M+1)\lambda_n}$ implies
\begin{equation}{\label{eq5.33}}\lambda_n=\frac{M}{2(M+1)\alpha_n}.
\end{equation}
 Now we only need to prove  the expressions of $\beta_{in}$ and
 $\sigma_{z_n}^2$.

(\ref{eq5.28}) and (\ref{eq5.30}) yield
$\lambda_n=\frac{M\beta_{in}\Sigma{n-1}}{(M+1)\sigma_{\mu}^2}$, thus
\begin{equation}{\label{eq5.34}}\beta_{in}=\frac{\lambda_n(M+1)\sigma_{\mu}^2}{M\Sigma_{n-1}}.
\end{equation}
Combining (\ref{eq5.30}) and (\ref{eq5.34}), we  get the expression
of $\sigma_{z_n}^2$ as
\begin{equation}{\label{eq5.35}}\sigma_{z_n}^2=\frac{1}{M}\left(\sigma_{\mu}^2-
\frac{\lambda_n^2(M+1)^2\sigma_{\mu}^4}{M\Sigma_{n-1}}\right).
\end{equation}

Next  we prove that the difference equation system consisted by
equation (\ref{eq5.25}), (\ref{eq5.26}), (\ref{eq5.32}),
(\ref{eq5.33}), (\ref{eq5.34}) and (\ref{eq5.35}) has a unique
solution.  We suppose that
$\lambda_n=\frac{\sqrt{\Sigma_{n-1}}}{(M+1)\sigma_{\mu}}a_n$, where
$a_n$ is a function of $n$. In turn, (\ref{eq5.32}) reduces to
\begin{equation}{\label{eq5.36}}\Sigma_{n}=\left(1-\frac{a_{n}^2}{M}\right)\Sigma_{n-1}.
\end{equation}
From this equation, we have $0<a_n<\sqrt{M}$  ($n=1,\cdots,N-1$),
and  $a_N=\sqrt{M}$. Then , it is easy to get
\begin{equation}{\label{eq5.37}}\alpha_n=\frac{M}{2(M+1)\lambda_n}=\frac{M\sigma_{\mu}}{2\sqrt{\Sigma_{n-1}a_n}},
\end{equation}
\begin{equation}{\label{eq5.38}}\beta_{in}=\frac{a_n\sigma_{\mu}}{M\sqrt{\Sigma_{n-1}}}.
\end{equation}
Then $(\ref{eq5.25})$  yields

\begin{equation}{\label{eq5.39}}
\frac{M}{2a_{n-1}\sqrt{\Sigma_{n-2}}}
=\frac{2(M+1)(1-M)a_n^2+(M-1)a_n^4+M^2(M+1)}{2M(M+1)a_n\sqrt{\Sigma_{n-1}}}.
\end{equation}
Substituting
$\Sigma_{n-1}=\left(1-\frac{a_{n-1}^2}{M}\right)\Sigma_{n-2}$ into
the above equation we  get
\begin{equation}{\label{eq5.40}}\frac{\sqrt{M(M-a_{n-1}^2)}}{2a_{n-1}}=
\frac{(1-M)a_n}{M}+\frac{(M-1)a_n^3}{2M(M+1)}+\frac{M}{2a_n}.
\end{equation}
Because $0<a_n<\sqrt{M}$($n=1,\cdots,N-1$), $a_N=\sqrt{M}$, the
above equation is equivalent to the following equation
\begin{equation}{\label{eq5.41}}a_{n-1}=\left(\frac{M^4(M+1)^2a_n^2}{M^3(M+1)^2a_n^2+
[2(M+1)(1-M)a_n^2+(M-1)a_n^4+M^2(M+1)]^2}\right)^{\frac{1}{2}}.
\end{equation}
From this equation we can know that every $a_n$  exists and is
unique for given $a_N=\sqrt{M}$. Therefore, the difference equation
system has  a unique solution.

{\bf {Proof of Proposition \ref{pro6}}}: By equation
$(\ref{eq5.41})$, we have
\begin{equation}{\label{eq6.1}}\frac{a_{n-1}^2}{a_n^2}=\frac{M^4(M+1)^2}{M^3(M+1)^2a_n^2+
[2(M+1)(1-M)a_n^2+(M-1)a_n^4+M^2(M+1)]^2}.
\end{equation}
Because
\begin{equation}\begin{aligned}{\label{eq6.2}}&M^3(M+1)^2a_n^2+
[2(M+1)(1-M)a_n^2+(M-1)a_n^4+M^2(M+1)]^2\\
=&M^4(M+1)^2+(M+1)^2(4M^2-3M^3)a_n^2+2M^2(M^2-1)a_n^4\\&+
[2(M+1)(1-M)a_n^2+(M-1)a_n^4]^2.
\end{aligned}
\end{equation}
Let $b_n=a_n^2$, we have
\begin{equation}{\label{eq6.3}}\frac{a_{n-1}^2}{a_n^2}=\frac{M^4(M+1)^2}{M^4(M+1)^2+F(b_n)},
\end{equation}
where $F(x)$ satisfies
\begin{equation}\begin{aligned}{\label{eq6.4}}F(x)=&x\left\{(M+1)^2(4M^2-3M^3)+2M^2(M^2-1)x+x
[2(M+1)(1-M)+(M-1)ax]^2\right\}\\
=&x\left\{(M-1)^2x^3-4(M+1)(M-1)^2x^2+(M^2-1)(6M^2-4)x+(M+1)^2(4M^2-3M^3)\right\}.
\end{aligned}
\end{equation}
It is easy to get that $F(M)=4M^2$.

Now, for any fixed $N$,  we prove  $\frac{a_{n-1}^2}{a_n^2}\leq 1$.
Let
\begin{equation}\begin{aligned}{\label{eq6.5}}f(x)\doteq
(M-1)^2x^3-4(M+1)(M-1)^2x^2+(M^2-1)(6M^2-4)x+(M+1)^2(4M^2-3M^3),
\end{aligned}
\end{equation}
so
\begin{equation}\begin{aligned}{\label{eq6.6}}f'(x)=
3(M-1)^2x^2-8(M+1)(M-1)^2x+(M^2-1)(6M^2-4).
\end{aligned}
\end{equation}
Let $f'(x)=0$, since the $\Delta$ of the above quadratic equation is
\begin{equation}\begin{aligned}{\label{eq6.7}}\Delta=&
64(M+1)^2(M-1)^4-12(M-1)^2(M^2-1)(6M^2-4)\\
&=8(M-1)^2(M^2-1)(-M^2-2)<0,
\end{aligned}
\end{equation}
so  for any $x$, $f(x)$ is increasing.  On the other hand,
$f(0)=(M+1)^2(4M^2-3M^3)<0$, $f(M)=4M>0$, so $f(x)$ have a unique
solution in $(0,M)$, we denote the solution is $A$.

 If for any $n,1\leq n \leq N-1$, $a_n^2\in [A,M)$, by
\begin{equation}\begin{aligned}{\label{eq6.8}}\frac{a_{n-1}^2}{a_n^2}=\frac{M^4(M+1)^2}{M^4(M+1)^2+a_n^2f(a_n^2)},
\end{aligned}
\end{equation}
  and $f(a_n^2)>0$, we have $\frac{a_{n-1}^2}{a_n^2}\leq1$, then the
  proposition is proved.

So we only need to prove that for  any $n,1\leq n \leq N-1$,
$a_n^2\in [A,M)$.  In order to prove this conclusion, we fist give
the following proposition.
\begin{proposition}\label{pro7} The real root of $f(x)=0$, denoted by $A$,  satisfies
\begin{equation}A\in\left(M-\frac{4M}{(M-1)(M^3+3M^2+4M-4)},
M-\frac{4M}{M^2(M+1)^2+4}\right).
\end{equation}
\end{proposition}
\begin{proof}: From $a_N=\sqrt{M}$ and equation $(\ref{eq5.41})$, we
have
\begin{equation}\begin{aligned}{\label{eq6.9}}a_{N-1}=\sqrt{\frac{M^3(M+1)^2}{M^2(M+1)^2+4}}.
\end{aligned}
\end{equation}
In order to get the range of $A$, we first compute
$f(M-\varepsilon)$, where $0<\varepsilon<M$.
\begin{equation}\begin{aligned}{\label{eq6.10}}f(M-\varepsilon)=&(M-1)^2(M-\varepsilon)^3
-4(M+1)(M-1)^2(M-\varepsilon)^2+\\&(M^2-1)(6M^2-4)(M-\varepsilon)+(M+1)^2(4M^2-3M^3)\\
=&-(M-1)^2\varepsilon^3-(M-1)^2(M+4)\varepsilon^2-\\&(M-1)(M^3+3M^2+4M-4)\varepsilon+4M,
\end{aligned}
\end{equation}
so from the above equation we have
\begin{equation}\begin{aligned}{\label{eq6.11}}f\left(M-\frac{4M}{(M-1)(M^3+3M^2+4M-4)}\right)<0,
\end{aligned}
\end{equation}
i.e.
\begin{equation}\begin{aligned}{\label{eq6.12}}A>M-\frac{4M}{(M-1)(M^3+3M^2+4M-4)}.
\end{aligned}
\end{equation}
On the other hand,  $(\ref{eq6.10})$ implies
\begin{equation}\begin{aligned}{\label{eq6.13}}&f\left(M-\frac{4M}{M^2(M+1)^2+4}\right)\\
=&\frac{1}{[M^2(M+1)^2+4]^3}\{-64M^3(M-1)^2-16M^2(M-1)^2(M+4)[M^2(M+1)^2+4]\\&+32M^2[M^2(M+1)^2+4]^2\}\\
=&\frac{32M^2}{[M^2(M+1)^2+4]}-\frac{16M^2(M-1)^2(M+4)}{[M^2(M+1)^2+4]^2}-
\frac{64M^3(M-1)^2}{[M^2(M+1)^2+4]^3},
\end{aligned}
\end{equation}
and
\begin{equation}\begin{aligned}{\label{eq6.14}}
&\frac{M^3(M+1)^2}{M^2(M+1)^2+4}f\left(M-\frac{4M}{M^2(M+1)^2+4}\right)\\
=&\frac{1}{[M^2(M+1)^2+4]^4}[M^7(M+1)^3(2M^3+M^2+2M-1)+\\&4M^5(4M^4+6M^3+4M^2+6M-4)]
>0,
\end{aligned}
\end{equation}
so we have
\begin{equation}\begin{aligned}{\label{eq6.16}}
A<M-\frac{4M}{M^2(M+1)^2+4}.
\end{aligned}
\end{equation}
\end{proof}
Next we use Proposition \ref{pro7} to prove that  $A\leq a_n^2<M$
implies $A\leq a_{n-1}^2<M$. We use the backward induction, using
$(\ref{eq6.8})$,
\begin{equation}\begin{aligned}{\label{eq6.17}}a_{n-1}^2=\frac{M^4(M+1)^2a_n^2}{M^4(M+1)^2+a_n^2f(a_n^2)},
\end{aligned}
\end{equation}
i.e.
\begin{equation}\begin{aligned}{\label{eq6.18}}A\leq\frac{M^4(M+1)^2a_n^2}{M^4(M+1)^2+a_n^2f(a_n^2)}<M.
\end{aligned}
\end{equation}
By the left of the above inequality, we have
\begin{equation}\begin{aligned}{\label{eq6.19}}M^4(M+1)^2A+a_n^2f(a_n^2)A-M^4(M+1)^2a_n^2\leq
0.
\end{aligned}
\end{equation}
Let
\begin{equation}\begin{aligned}{\label{eq6.20}}g(x)\doteq&
M^4(M+1)^2A+xf(x)A-M^4(M+1)^2x\\
=&M^4(M+1)^2A+AF(x)-M^4(M+1)^2x.
\end{aligned}
\end{equation}
It is easy to get that $g(0)=M^4(M+1)^2A>0$, $g(A)=0$. By
Proposition $\ref{pro6}$, we get
\begin{equation}\begin{aligned}{\label{eq6.21}}g(M)=&
M^4(M+1)^2A+Mf(M)A-M^4(M+1)^2M\\
=&A[M^4(M+1)^2+4M^2]-M^4(M+1)^2M<0.
\end{aligned}
\end{equation}
So if we can prove that $g(x)$ is decreasing in $[0,M)$, we can get
the conclusion that
\begin{equation}\begin{aligned}{\label{eq6.22}}A\leq
a_n^2<M\Leftrightarrow (\ref{eq6.19}).
\end{aligned}
\end{equation}
Next we prove $g(x)$ is decreasing in $[0,M)$. Since
\begin{equation}\begin{aligned}{\label{eq6.23}}g'(x)=AF'(x)-M^4(M+1)^2,
\end{aligned}
\end{equation}
from equation $(\ref{eq6.4})$, we have
\begin{equation}\begin{aligned}{\label{eq6.24}}F'(x)=&4(M-1)^2x^3-
12(M+1)(M-1)^2x^2+2(M^2-1)(6M^2-4)x+\\&(M+1)^2(4M^2-3M^3),
\end{aligned}
\end{equation}
\begin{equation}\begin{aligned}{\label{eq6.25}}F''(x)=12(M-1)^2x^2-
24(M+1)(M-1)^2x+2(M^2-1)(6M^2-4).
\end{aligned}
\end{equation}
Let $F''(x)=0$, since the $\Delta$ of the  quadratic equation is
\begin{equation}\begin{aligned}{\label{eq6.26}}\Delta=&24\times24(M+1)^2(M-1)^4-96(M-1)^2(M^2-1)(6M^2-4)\\
=&96(M-1)^2[6(M+1)^2(M-1)^2-(M^2-1)(6M^2-4)]\\
=&-192(M-1)^2(M^2-1)<0.
\end{aligned}
\end{equation}
So for all $x\in[0,M)$, $F''(x)>0$, i.e. $F'(x)$ is a strictly
increasing function in $[0,M)$. By equation $(\ref{eq6.24})$ we
have,
\begin{equation}\begin{aligned}{\label{eq6.27}}F'(M)=&
4(M-1)^2M^3-12(M+1)(M-1)^2M^2+2(M^2-1)(6M^2-4)M+\\&(M+1)^2(4M^2-3M^3)\\
=&M^5+2M^4+M^3-8M^2+8M.
\end{aligned}
\end{equation}
So
\begin{equation}\begin{aligned}{\label{eq6.27}}g'(x)=&AF'(x)-M^4(M+1)^2\\
<&AF'(M)-M^4(M+1)^2\\
<&M^6+2M^5+M^4-8M^3+8M^2-M^6-2M^5-M^4\\
=&-8M^3+8M^2<0.
\end{aligned}
\end{equation}
That means  $g(x)$ is decreasing in $[0,M)$. Then we prove
Proposition \ref{pro6}.

\textbf{Proof of Theorem $\ref{th5}$:} Given $t\in (0,1)$, choose
$N_1<N_2<\cdots<N_j<\cdots \rightarrow +\infty$ satisfy
$2+(1-t)N_j\leq (1-t)N_{j+1}<N_j-1$, $j=1,2,\cdots$. For every
$N_j$, let $n(N_j)=[tN_j]$, $j=1,2,\cdots.$ By $N_{j+1}>N_j$ and
$(1-t)(N_{j+1}-N_j)\geq 2$, it is easy to get that $n(N_{j+1})>
n(N_{j})$, $n(N_{j+1})> N_{j+1}-N_j$. For the market having $N_j$
auctions,  we denote $a_m^{(N_j)}(m=1,2,\cdots,N_j)$ the sequence
defined by $(\ref{eq5.41})$.
 So
$a_{N_j}^{(N_j)}=a_{N_{j+1}}^{(N_{j+1})}=\sqrt{M}$. By
$(\ref{eq5.41})$, it is easy to get:
$$a_{N_j-l}^{(N_j)}=a_{N_{j+1}-l}^{(N_{j+1})}, ~~~~l=0,1,\cdots,N_j.$$
So $$a_k^{(N_{j+1})}=a_{k-(N_{j+1}-N_j)}^{(N_j)},~~~~ (\forall
k=N_{j+1}-N_j,\cdots,N_{j+1}),$$  and
$$a_{n(N_{j+1})}^{(N_{j+1})}=a_{n(N_{j+1})-(N_{j+1}-N_j)}^{(N_{j})}.$$
Moreover, $(1-t)N_{j+1}<N_j-1$ implies  $n(N_j)\geq
n(N_{j+1})-(N_{j+1}-N_j).$  So,  according to Proposition
$\ref{pro7}$ in Section $5$ and Proposition $\ref{pro6}$ we know
that, $A\leq a_{n(N_{j+1})}^{(N_{j+1})}\leq a_{n(N_j)}^{(N_j)}.$
Hence,  when $j\rightarrow \infty$ the limit of $a_{n(N_j)}^{(N_j)}$
exists, and denote it by $a$. Furthermore,by $a_{n(N_j)}^{(N_j)}\leq
a_{n(N_j)-1}^{(N_j)}=a_{n(N_j)-(N_j-N_{j-1})}^{(N_{j-1})}\leq
a_{n(N_{j-1})}^{(N_{j-1})}$, we know that $a_{n(N_j)-1}^{(N_j)}$
also converges to $a$ as $j\rightarrow +\infty.$ By $(\ref{eq6.8})$,
$a^2$ is just the real root of $f(x)=0$ in $[A,M)$, i.e. $a^2=A,$
where $f(x)$ is defined by (\ref{eq6.5}).

(i) When $M=1$, $(\ref{eq5.41})$ is reduce to
$a_{n-1}=\sqrt{\frac{a_n^2}{a_n^2+1}}$ and $a_N=1$. It is easy to
get, for any fixed $N_j$, and $n=1,\cdots,N_j$,
$a_n^2=\frac{1}{N_j-n+1}$.  By $(\ref{eq5.36})$,
$$\Sigma_n=\frac{N_j-n}{N_j-n+1}\Sigma_{n-1}=\frac{N_j-n}{N_j}\Sigma_0$$
i.e. $\Sigma_n-\Sigma_{n-1}=-\frac{\Sigma_0}{N_j}.$ It is just the
results of Proposition $4$ in Hudart, Hughes and Levine (2001).
Since
$$\Sigma_{n(N_j)}^{(N_j)}=\frac{N_j-n(N_j)}{N_j}\Sigma_0=\frac{N_j-[tN_j]}{N_j}\Sigma_0,$$
$\Sigma_{n(N_j)}^{(N_j)}$ converges to $\Sigma_t$ when $j\rightarrow
\infty,$ where $\Sigma_t=(1-t)\Sigma_0.$  By
$\lambda_n=\frac{\sqrt{\Sigma_{n-1}}a_n}{(M+1)\sigma_{\mu}},$ we get
$\lambda_{n(N_j)}^{(N_j)}=\sqrt{\frac{\Sigma_0}{N_j}}\rightarrow 0,$
when $j\rightarrow+\infty$. So, $\lambda_t=0.$ Combing
$\Sigma_{n(N_j)-1}^{(N_j)}=\frac{N_j-n(N_j)+1}{N_j}\Sigma_0=\frac{N_j-[tN_j]+1}{N_j}\Sigma_0$
and $(\ref{eq5.34})$, we get $\beta_{it}=0.$

(ii) When $M\geq 2,$ we will first prove that
$\Sigma_{n(N_k)}^{(N_k)}$ and $\Sigma_{n(N_k)-1}^{(N_k)}$ converge
to the same limit, and we denote it as $\Sigma_t.$ Since  when
$k\rightarrow+\infty$ , $a_{n(N_k)}^{(N_k)}\rightarrow \sqrt{A}$
 and $\Sigma_{n(N_k)}^{(N_k)}$ satisfy
\begin{equation}\begin{aligned}{\label{eq6.30}}\Sigma_{n(N_k)}^{(N_k)}
=\left(1-\frac{{a_{n(N_k)}^{(N_k)}}^2}{M}\right)\Sigma_{n(N_k)-1}^{(N_k)},
\end{aligned}
\end{equation}
we have
\begin{equation}\begin{aligned}{\label{eq6.31}}\Sigma_{n(N_k)}^{(N_k)}
&=\prod_{i=1}^{n(N_k)}\left(1-\frac{{a_{i}^{(N_k)}}^2}{M}\right)\Sigma_0\\
&\leq\left(1-\frac{{a_{1}^{(N_k)}}^2}{M}\right)^{[tN_k]}\Sigma_0\\
&\leq \left(1-\frac{A}{M}\right)^{[tN_k]}\Sigma_0.
\end{aligned}
\end{equation}
In the same way, by ${a_{n(N_k)}^{(N_k)}}^2\to A<M$, we have
\begin{equation}\begin{aligned}{\label{eq6.33}}\Sigma_{n(N_k)}^{(N_k)}
&=\prod_{i=1}^{n(N_k)}\left(1-\frac{{a_{i}^{(N_k)}}^2}{M}\right)\Sigma_0\\
&\geq\left(1-\frac{{a_{n(N_k)}^{(N_k)}}^2}{M}\right)^{[tN_k]}\Sigma_0
\\&\geq\left(1-\frac{A}{M}+o(1)\right)^{[tN_k]}\Sigma_0\to 0.
\end{aligned}
\end{equation}
Combing (\ref{eq6.31}) and (\ref{eq6.33}) we have the (\ref{eq**}),
and when $j\rightarrow+\infty,$ $\Sigma_{n(N_k)}^{(N_k)}\rightarrow
\Sigma_t=0$. Combing with (\ref{eq6.30}), we have, when $j
\rightarrow +\infty$, $\Sigma_{n(N_{j})-1}^{(N_{j})}$ also converges
to $\Sigma_t.$

By
$$\lambda_{n(N_j)}^{(N_j)}=\frac{\sqrt{\Sigma_{n(N_{j})-1}^{(N_{j})}}}{2\sigma_{\mu}}a_{n(N_{j})}^{(N_{j})}$$
we get that $\lambda_t=0$,
$\alpha_t=\frac{M}{2(M+1)\lambda_t}=+\infty.$ And it is easy to get
$\delta_t=\gamma_t=0$, $\beta_{it}=+\infty$
$\sigma_{z_t}^2=\frac{\sigma_{\mu}^2}{M}(1-\frac{A}{M}).$ Then
Theorem $\ref{th5}$ is proved.

\makeatletter
\renewcommand{\@biblabel}[1]{}
\renewcommand{\@cite}[1]{ }
\makeatother

\end{document}